# Spatiotemporal Optical Vortices From All-Dielectric Bilayer Metagratings

Ken Qin[1], Shijie Kang[1], Aoning Luo[1], Yiyi Yao[1], Xiexuan Zhang[1], Hanchuan Chen[1], Yahan Xiao[2], Yangsong Ye[3], Junqing Shi[1], Xusheng Xia[4,*], Haitao Li[1,*], and Xiaoxiao Wu[1,*]

[1]*Modern Matter Laboratory and Advanced Materials Thrust, The Hong Kong University of Science and Technology (Guangzhou), Nansha, Guangzhou 511400, Guangdong, China*

[2]*School of Integrated Circuit Science and Engineering, University of Electronic Science and Technology of China, Chengdu 610054, China*

[3]*Center for Information Photonics and Energy Materials, Shenzhen Institute of Advanced Technology, Chinese Academy of Sciences, Shenzhen 518055, China*

[4]*School of Physics and Mechanics, Wuhan University of Technology, Wuhan 430070, China*

[*]*To whom correspondence should be addressed. E-mails: xsxia@whut.edu.cn (X. Xia), haitaoli@hkust-gz.edu.cn (H. Li) and xiaoxiaowu@hkust-gz.edu.cn (X. Wu)*

## Abstract

Spatiotemporal optical vortices (STOVs) carry transverse orbital angular momentum within the space-time domain, rendering them powerful tools for constructing high-dimensional and quantum optical fields. However, most existing approaches rely on highly lossy metallic structures or complex pulse-shaping systems. Here, we propose an all-dielectric route to STOV generation based on symmetry-protected bound states in the continuum (BICs) in a bilayer metagrating and provide proof-of-concept validation of its key momentum-frequency signatures. By simply introducing a lateral shift between the upper and lower layers of the vertical slots on the dielectric metagrating, the Γ-point BIC transforms into a quasi-BIC (qBIC) with momentum-dependent directional radiation and asymmetric coupling. This qBIC further leads to an isolated zero-transmission dip associated with a clear phase singularity and branch cut in the momentum-frequency response, enabling stable STOV generation under excitation by a spatiotemporal Gaussian pulse. The multipole analysis of the qBIC reveals the key role of the Kerker-type interference between its electric and magnetic dipoles and associated asymmetric coupling in the STOV generation. Experimentally, free-space transmission measurements reveal a transmission zero and a branch cut that agree excellently with theoretical analysis. Therefore, our work provides an experimentally validated, scalable route for manipulating spatiotemporal optical fields on low-loss all-dielectric metasurfaces via only gliding offsets, with potential applications in directional coupling of quantum light sources and spatiotemporal shaping of single-photon wave packets.

**Keywords:** spatiotemporal optical vortex (STOV); quasi-bound states in the continuum (qBICs); all-dielectric; bilayer metagrating; Kerker effect.

## I. INTRODUCTION

Vortices are ubiquitous in wave physics and are characterized by phase singularities accompanied by vanishing field amplitudes at the cores.[1,2] In optics, such vortex states—referred to as optical vortices (OVs)—carry well-defined orbital angular momentum (OAM)[3,4] and have enabled a wide range of applications, including optical manipulation,[5,6] structured light communication,[7,8] and high-resolution imaging.[9,10] To date, most studies have focused on longitudinal OAM beams, whose angular-momentum vectors are aligned with the propagation directions.[11–14] By contrast, transverse OAM beams, in which the OAM is oriented perpendicular to the propagation axis,[15–18] represent a distinct class of structured light fields.

A representative class of OVs carrying transverse OAM is spatiotemporal optical vortices (STOVs) that host phase singularities jointly within the space-time domain.[19–21] They hold strong promise for group-delay engineering,[22,23] quantum wave-packet shaping,[24,25] and high-dimensional quantum encoding.[26–28] Yet controllable STOV generation has largely depended on Fourier-optics pulse shapers that imprint spiral phases in the frequency-space domain, requiring bulky and complex setups[29–31] that are difficult to integrate on chip. All-dielectric metasurfaces represent a compact and stable approach,[32] especially for quantum photonics where low loss is critical: high-index dielectric structures preserve strong scattering and design freedom while suppressing Ohmic loss.[33] A kind of all-dielectric spatiotemporal differentiator can generate STOVs within an ultrathin device,[34] but the asymmetric nanostructures are difficult to fabricate with high yield using standard lithography and etching. Moreover, surface roughness, etch-induced damage, and nonuniformity introduce additional loss, reducing single-photon count rates and coincidence rates. Meanwhile, all-dielectric metasurfaces support symmetry-protected bound states in the continuum (SP-BICs)[35] and their symmetry-broken quasi-BIC (qBIC) counterparts naturally support high quality factors (Q-factors) and subwavelength field localization. These characteristics enable a robust scattering channel that can destructively interfere with the background transmission, resulting in a zero transmission dip as the resonance singularity.[36] However, existing mirror-symmetry-breaking implementations, such as slanted metagratings,[37] face a practical scalability barrier for quantum photonics: maintaining a uniform, well-calibrated sidewall tilt and low-scattering interfaces over millimeter- to wafer-scale areas is challenging. Such large-scale fabrication can easily introduce spatially varying symmetry breaking, thereby shifting or weakening the targeted singularity conditions, leading to substantial loss and phase noise that compromise STOV mode purity and quantum-state fidelity. In contrast, breaking symmetry via a dislocation between stacked gratings provides a more fabrication-friendly route. Recently, in metallic bilayer-grating platforms, dislocation-based symmetry breaking has been successfully used to couple free-space excitation into unidirectional spoof surface plasmon polariton (SSPP) modes or to produce Fano-type resonant responses.[38,39] This, in turn, yields a spatiotemporal transfer function featuring a phase singularity with a 2π winding in the momentum-frequency domain, enabling the generation of STOVs. However, it remains unclear how these results can be generalized into a predictive, design-oriented mechanism that establishes a mapping from the momentum-frequency response to the space-time STOV waveform, particularly for scalable, low-loss dielectric metagratings where the SSPP-based picture is not directly applicable.

Here, we propose a proof-of-principle, qBIC-based route to STOV generation through all-dielectric bilayer metagratings as shown in Figure 1(a). A dislocation between stacked slots breaks the mirror symmetry and converts an SP-BIC into a qBIC with momentum-dependent asymmetric radiative coupling. This gives rise to a spiral 2π phase winding around a singularity in the Bloch-wavenumber-frequency ($k_x$ - $f$) parameter space. Under normal-incidence excitation with a spatiotemporal Gaussian pulse, our analysis reveals a clear central null and spiral phase for the transmitted pulse in the periodic direction and time ($x$ - $T$) domain, confirming STOV formation. Furthermore, multipole analysis shows that symmetry breaking induces the orthogonal electric dipole and magnetic dipole,[40] whose Kerker-type interference results in the directional and asymmetric radiation.[41–43] Building on this physical picture, the Lorentz model[44] we developed shows that both the transmission zero for various symmetry-breaking parameters and the associated phase branch cut stem from the underlying zero-pole structure in the complex-frequency plane. Experimental measurements in oblique-incidence angle and frequency space ($\theta$ - $f$) further validate the predicted amplitude and phase behavior, showing excellent agreement with theory and simulations. Overall, this work demonstrates qBIC-driven STOV generation on a low-loss dielectric platform largely independent of materials and provides a transferable framework for directional quantum-light coupling, spatiotemporal wave-packet shaping, and high-dimensional photonic encoding.

## II. RESULTS

### A. BIC in symmetric bilayer metagrating

Figure 1(b) shows a side view of the bilayer metagrating: the metagrating has a period $a$ = 3 mm along the $x$ direction, a total thickness $h_1$ = 1.5 mm and upper and lower air-slot widths $d$ = 1 mm each with depth $h_2$ = $h_3$ = 0.6 mm. The dielectric material has a refractive index of $n$ = 3.13. While keeping the upper air-slot positions fixed, the lower air slots are shifted along the $x$ direction by a distance $\delta d$, selectively breaking the mirror symmetry of the structure. The band structure of the metagrating when $\delta d$ = 0 mm (symmetric case) is shown in Figure 1(c). The blue solid curve corresponds to the transverse electric (TE) mode, whose electric field is polarized in the $x$ - $z$ plane. It constitutes the resonance branch of interest in this work. The red dashed curve denotes the transverse magnetic (TM) mode, whose electric field is polarized along the $y$ direction. Figure 1(d) plots the Q-factor of the TE mode as a function of $k_x$. The Q-factor diverges at the Γ point, indicating that radiative loss is completely suppressed in the symmetric limit, forming an ideal BIC. Figures 1(e) and 1(f) show the distributions of transmission amplitude and phase of the TE mode, respectively, in the $k_x$ - $f$ space. At the Γ point, the transmission amplitude exhibits a distinct "dark spot" corresponding to an SP-BIC that cannot be excited by an incident plane wave. The transmission phase is also perfectly symmetric with respect to $k_x = 0$, as is clearly visible in Figure 1(f).

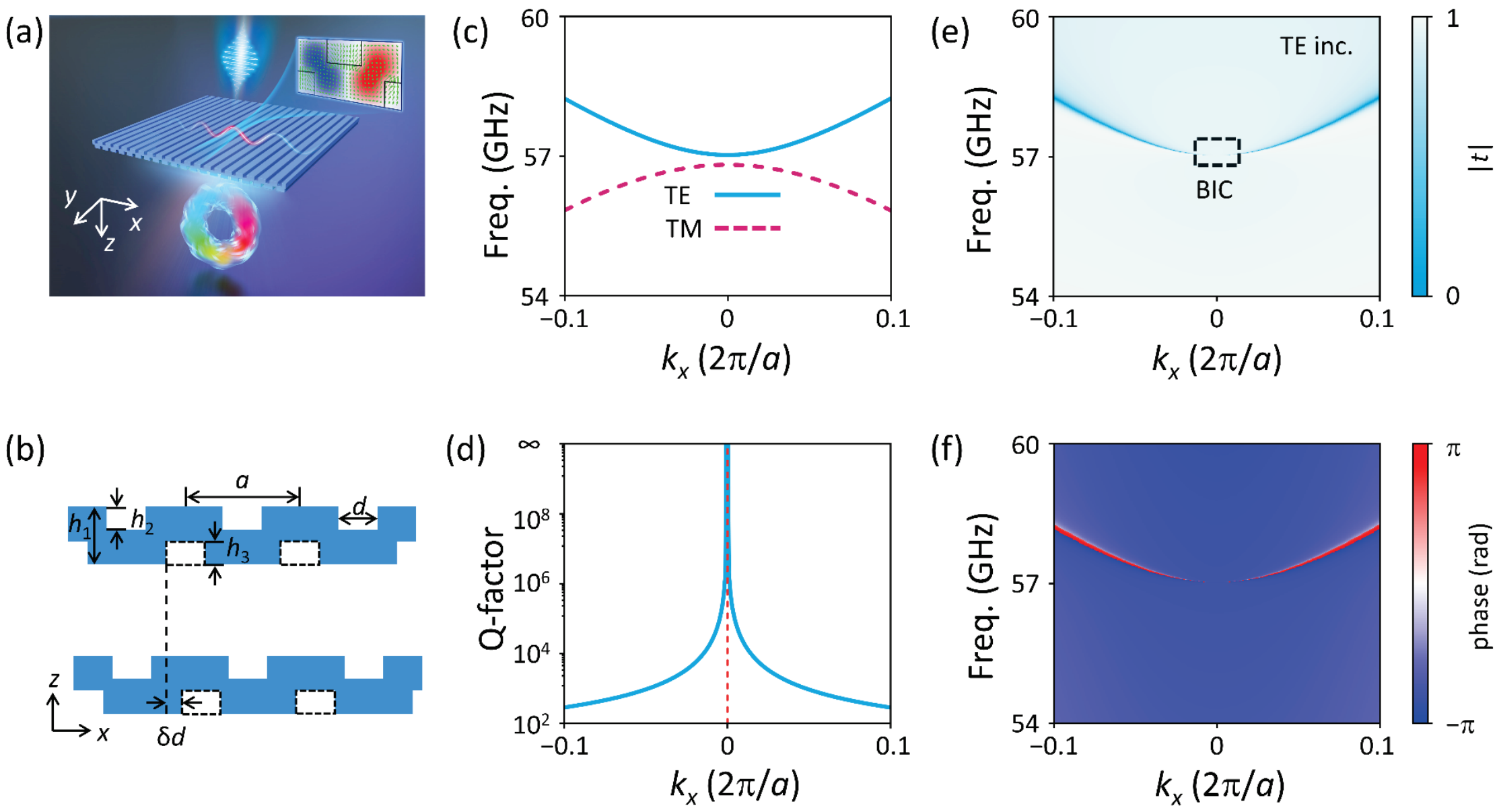

**Figure 1. All-dielectric bilayer metagrating and BIC.** (a) Schematic illustration of generating an STOV using all-dielectric bilayer metagrating under excitation by a spatiotemporal Gaussian beam. (b) Side view of the bilayer metagrating. Blue regions represent dielectric material with a refractive index $n = 3.13$, and white regions represent air slots. The structure period is $a$, thickness $h_1$, upper and lower air-slot width $d$, and depth $h_2$ and $h_3$. The upper panel corresponds to the mirror-symmetric case; the lower panel shows the symmetry-broken structure after shifting the lower air slots by $\delta d$ along the $x$ direction. (c) Band structure under preserved mirror symmetry. The blue solid line denotes the TE mode; the red dashed line denotes the TM mode. (d) Q-factor of the TE mode as a function of $k_x$ with the mirror symmetry. (e) and (f) The amplitude and phase maps of transmission coefficient in the $k_x$ - $f$ space under normal TE incidence.

## B. qBIC and STOVs

When a finite gliding is introduced into the structure to offset the slots, the mirror symmetry of the bilayer metagrating is broken. The SP-BIC degrades into a radiative qBIC. Figure 2(a) shows the normal-incidence transmission spectra for different asymmetry parameters $\delta d$. As $\delta d$ increases, the resonance linewidth of the transmission dip gradually broadens, indicating increased radiative loss and a reduced Q-factor for the qBIC. A quantitative analysis of the radiative quality factor reveals an inverse quadratic dependence on the asymmetry parameter, $Q \propto \delta d^{-2}$, which is a hallmark of SP-BICs perturbed by structural asymmetry[36] (see Figure S1 in the Supplementary Materials). Nevertheless, in the lossless scenarios across all such asymmetry parameters, the transmission dip due to the qBIC remains zero for normal incidence, ensuring existence and persistence of the transmission-zero singularity, as will be discussed in Section II C.

Beyond normal incidence, we plot the transmission amplitude and phase distributions in the $k_x$ - $f$ space. We confirm that a transmission zero emerges at the Γ point (Figure 2(b)), and non-zero transmission arises when deviating from the Γ point. While the transmission amplitude is symmetric with respect to $k_x = 0$, we find that a distinct branch cut appears on the $+k_x$ side of the phase map (Figure 2(c)). This branch cut is accompanied by a phase singularity: the phase winds by $2\pi$ along a clockwise loop encircling the zero-transmission singular point, indicating a winding number of −1. Owing to the Fourier correspondence between the $k_x$ - $f$ and $x$ - $T$ domains, such a momentum-frequency-domain signature is expected to manifest in the space-time domain as a central-null amplitude accompanied by a spiral phase.

To verify this correspondence, we perform FDTD simulations for a normally incident pulse with a spatiotemporal Gaussian envelope. The pulse is launched from a fixed source plane and propagates normally along the $+z$ direction. At the source plane, the input pulse, specifically, its magnetic field, is written as:

$$H_y(x,T) = H_0 \exp\left[-\frac{(x-x_0)^2}{w_x^2}\right] \exp\left[-\frac{(T-T_0)^2}{\tau^2}\right] \cos(\omega_0 T) \tag{1}$$

where $w_x$ is the beam waist radius, $x_0$ is the spatial center position, $\tau$ is the pulse duration, $T$ denotes time, $T_0$ is the temporal center, and $\omega_0$ is the central angular frequency.

Figures 2(d)-(f) show the magnitude and phase evolution of the transmitted field (represented by the $H_y$ field) in the $x$ - $T$ space under normal incidence, with $w_x = 45$ mm, $\delta d = 0.1, 0.2, 0.3$ mm, and the corresponding pulse durations $\tau = 10, 6, 3$ ns. To extract the transmitted complex-envelope field in the $x$ - $T$ domain, we apply a Hilbert transform[45,46] to the simulated time-domain signal to reconstruct the analytic (complex) field. In all three cases, the output exhibits a characteristic vortex-like spatiotemporal interference pattern (upper panel) with a spiral phase (lower panel), consistent with the phase branch-cut behavior in Figure 2(c), and the observed results from the Hilbert transform confirm the winding number of −1 for the generated STOV. The physical origin of the STOV formation can be further elucidated by the instantaneous field distribution in the $x$ - $z$ plane, which reveals a fork-like interference pattern and an associated phase singularity (see Figure S2 in the Supplementary Materials).

As $\delta d$ is further increased to around 0.4 mm, the associated phase vortices disappear because of mode hybridization, thereby suppressing STOV generation at large asymmetry (see Figure S3 in the Supplementary Materials). Importantly, when the lower air slots shift to the opposite direction ($\delta d < 0$), the transmission zero and phase singularity are preserved, but the phase winding direction is reversed, yielding a STOV with a winding number of +1 (see Figure S4 in the Supplementary Materials). In addition, as demonstrated in the Supplementary Materials Figure S5, cascading multiple asymmetric bilayer metagratings enables the generation of higher-order

STOVs through the accumulation of phase winding, further extending the flexibility of the proposed design.

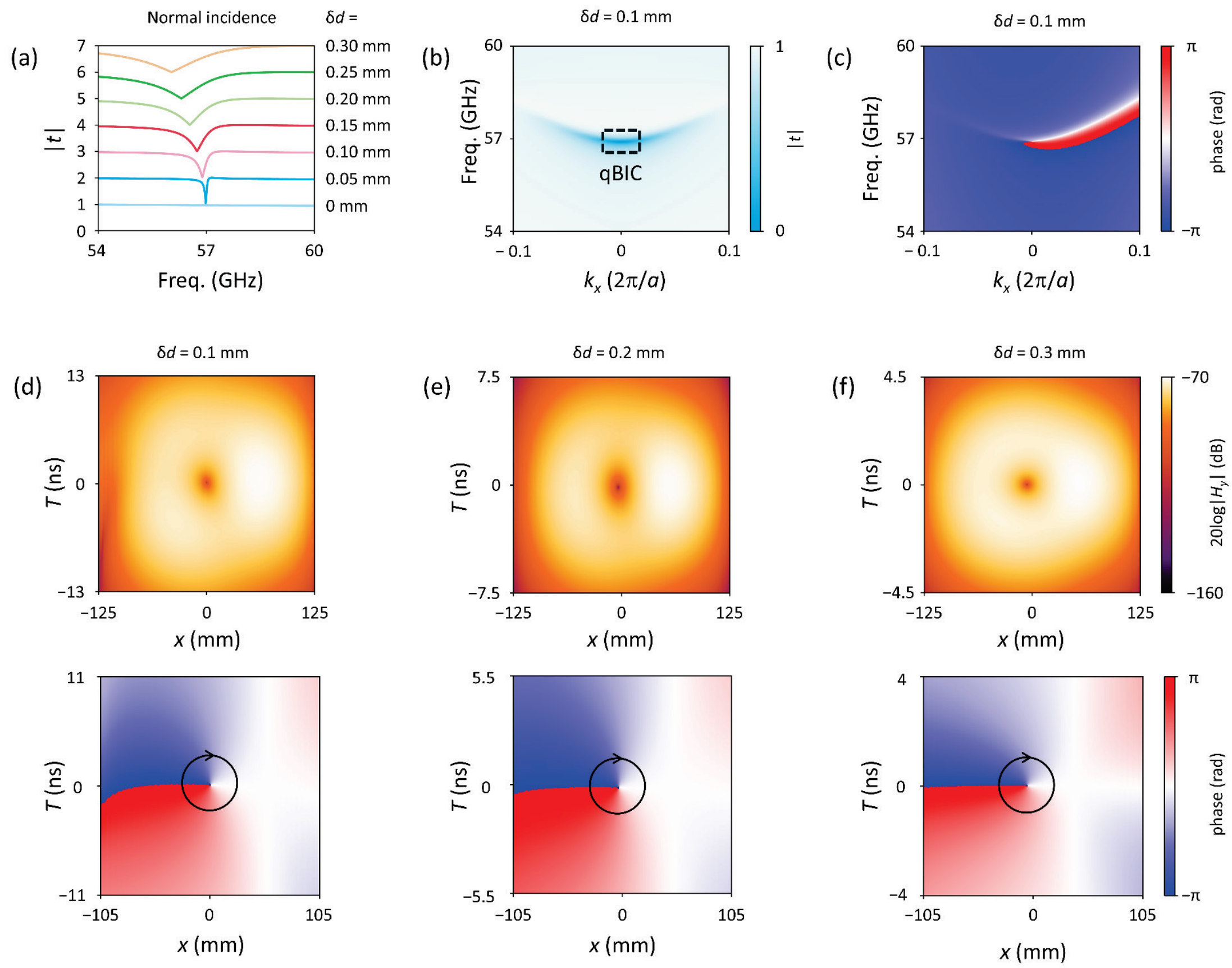

**Figure 2. Evolution of STOVs with structural asymmetry.** (a) Effect of different values for the asymmetry parameter $\delta d$ on the transmission spectrum $|t|$ under normal incidence. The spectra are vertically offset by unity in sequence for clarity. (b) and (c) Amplitude and phase distributions of the transmission coefficient in the $k_x$ - $f$ space for $\delta d$ = 0.1 mm. (d) - (f) Spatiotemporal evolution of the magnitude and phase of generated STOVs in the $x$ - $T$ space, obtained through simulations of the bilayer metagrating under excitation by a spatiotemporal Gaussian pulse. The beam waist is $w_x$ = 45 mm, and the asymmetry parameters are $\delta d$ = 0.1, 0.2, 0.3 mm, respectively, with corresponding pulse durations $\tau$ = 10, 6, 3 ns.

### C. Multipolar interference and directional radiation

To further elucidate the radiation mechanism of the qBIC mode after the structural mirror symmetry is broken, we first perform a multipole decomposition of the radiation fields of the eigenmodes of the structure. Figure 3(a) shows the scattering power of different multipole channels as a function of $k_x$ when $\delta d$ = 0 mm. It can be seen that, near $k_x$ = 0, the $z$-component of the electric dipole $ED_z$ and the $z$-component of the toroidal dipole $TD_z$ remain dominant and vary only weakly with $k_x$, whereas the radiation powers of the $x$-component of the electric dipole $ED_x$ and $y$-component of the magnetic dipole $MD_y$ are strongly suppressed at the Γ point. When a finite slot shift $\delta d$ is applied (Figure 3(b)), $ED_x$ and $MD_y$ increase significantly over the entire $k_x$ axis and exhibit strong $k_x$-dependence.

Figures 3(c) and 3(d) provide a more intuitive visualization of these multipolar configurations at the Γ point. In the upper schematic, white arrows indicate the magnitude and direction of the magnetic field $H_y$ at the two ends of a unit cell (one period of the metagrating), and the corresponding $TD_z$ and $MD_y$ are labeled with black symbols. When $\delta d$ = 0 mm, $H_y$ at the two ends has equal magnitudes but opposite directions, yielding a spatially antisymmetric magnetic-field distribution within the unit cell. This distribution indicates circulating magnetic dipoles, manifesting as a pronounced $TD_z$ and an almost vanishing net $MD_y$. After symmetry breaking, although the two ends still carry oppositely directed magnetic fields, their magnitudes become unequal. Consequently, in addition to the original toroidal dipole contribution, a non-zero magnetic dipole $MD_y$ emerges. The lower 3D field maps show the $H_y$ distribution in the plane $y$ = 0, with red arrows marking the local magnetic-field direction and magnitude, visually confirming the asymmetric magnetic-field distribution and coexistence of toroidal and magnetic dipoles illustrated in the schematic above when mirror symmetry breaks.

Away from the Γ point, for different values of $k_x$, the relative amplitudes and phases of $ED_x$ and $MD_y$ vary. Their interference in the far field, specifically a Kerker-type interference, drives a pronounced imbalance between the downward and upward radiation for non-zero $k_x$, which is crucial for a transmission zero even in the presence of material loss, as will be discussed later (Section II E). To visualize this effect, Figure 3(e) shows the $H_y$ field distributions in the upper and lower half-spaces for $\delta d$ = 0.1 mm at the Γ and off-Γ points, with $k_x = -0.05$ ($2\pi/a$), 0 and +0.05 ($2\pi/a$) as representative examples. When $k_x = -0.05$ ($2\pi/a$), the mode radiates more strongly into the lower half-space than into the upper half-space of the metagrating, whereas the situation reverses when $k_x$ = +0.05 ($2\pi/a$). Specifically, when $k_x$ = 0, the Poynting-vector fluxes radiated into the lower ($I_-$) and upper ($I_+$) half-spaces are equal. This is guaranteed by the identical air slot depth on the upper and lower layers that ensures a two-fold ($C_2$) rotational symmetry in the $x$ - $z$ plane (see Figure S6 in the Supplementary Materials). Figure 3(f) further quantifies this asymmetric radiation by plotting the ratio $I_-/I_+$ as a function of $k_x$. With a logarithmic scale of the flux ratio, the monotonic plot reveals that it spans several orders of magnitude and exhibits an exponentially asymmetric behavior depending on $k_x$. Such strongly directional emission arises from the Kerker-type interference between $ED_x$ and $MD_y$, which redistributes radiation preferentially into one half-space. The detailed far-field radiation patterns exhibiting the corresponding angular distributions are shown in Figure S7 of the Supplementary Materials. Moreover, the slope of $\log(I_-/I_+)$ is directly controlled by the structural asymmetry parameter $\delta d$, indicating that the degree of directional radiation can be continuously tuned via symmetry breaking, as further detailed in Figure S8 of the Supplementary Materials.

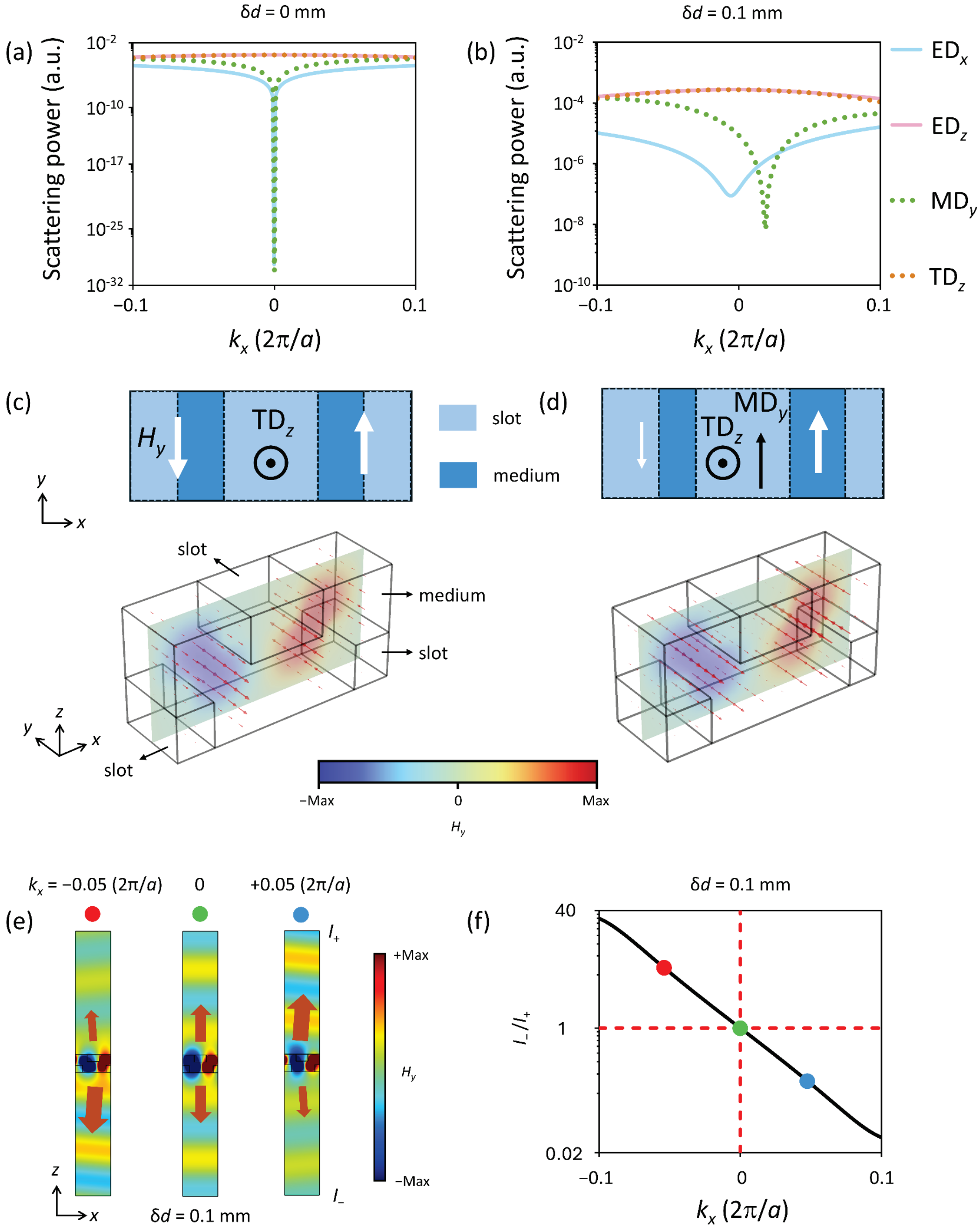

**Figure 3. Directional radiation and its multipolar interference origin.** (a) - (b) Multipole decomposition of the radiation of the eigenmodes for δ$d$ = 0 mm and δ$d$ = 0.1 mm. The blue solid line, pink solid line, green dashed line, and orange dashed line correspond to $ED_x$, $ED_z$, $MD_y$ and $TD_z$, respectively. (c) - (d) Schematics (upper panels) and 3D magnetic-field distributions (lower panels) of the symmetric structure (δ$d$ = 0 mm) and asymmetric structure (δ$d$ = 0.1 mm). In the schematics, white arrows indicate the magnitude and direction of the magnetic field $H_y$, and black symbols mark the directional configurations of the multipoles $TD_z$ and $MD_y$. The lower panels show the distribution of $H_y$ in the $y$ = 0 plane, with red arrows indicating the local magnitude and direction of the magnetic field. (e) Distributions of $H_y$ radiated into the upper and lower half-spaces for δ$d$ = 0.1 mm at $k_x$ = −0.05 (2π/$a$), 0 and +0.05 (2π/$a$). The size and direction of the arrows indicate the magnitude and direction of the radiation. (f) For δ$d$ = 0.1 mm, the ratio of the Poynting-vector flux in the lower half-space ($I_-$) to that in the upper half-space ($I_+$) as a function of $k_x$. Red, green and blue dots represent $k_x = -0.05$ ($2\pi/a$), 0 and +0.05 ($2\pi/a$) that correspond to the dots in Figure 3 (e), respectively.

## D. Lorentz oscillator model and phase singularities

Building on the asymmetry of downward and upward radiation about $\pm k_x$, we note that, within the operating frequency band, the leakage of a waveguide mode can be interpreted as the coherent superposition of a narrowband resonant channel dominated by the qBIC mode and a smooth background channel. Accordingly, the transmission coefficient can be analytically approximated using a model with an effective Lorentz oscillator plus background scattering:

$$t(k_x, f) = A(k_x) + \frac{B(k_x)\gamma_0 f_0}{f^2 - f_0^2 - 2if\gamma_0} \quad (2)$$

where $A(k_x)$ is the complex amplitude describing the non-resonant background channel, $B(k_x)$ is the complex coupling coefficient of the effective Lorentz oscillator, and $f_0$ and $\gamma_0$ correspond to the resonance frequency and the radiative linewidth, respectively. Figure 4(a) shows the fitting of this model to the full-wave transmission spectra under illumination from the lower side of the structure at $k_x = \pm 0.05$ ($2\pi/a$). The effective Lorentz model accurately reproduces the narrow resonance dip in amplitude and captures the rapid phase variation near resonance, confirming that the transmission response is well described by a single qBIC resonance channel within the relevant momentum range.

To understand the origin of the phase singularities, we analytically continue the transmission coefficient into the complex-frequency (Re($f$) - Im($f$)) plane and plot its phase distribution, as shown in Figures 4(b) and 4(c). When $k_x < 0$ (Figure 4(b)), the zero and the pole points both lie above the real-frequency axis, so the real-frequency sweep does not cross the branch cut and the transmission phase exhibits only a conventional small jump. This corresponds to the absence of a clear branch cut in the $k_x < 0$ region. In contrast, when $k_x > 0$ (Figure 4(c)), the pole point remains essentially fixed, but the zero point moves across the real-frequency axis into the opposite half-plane. The sweep along the real axis now effectively passes between the zero and the pole points, giving rise to an almost 2π topological phase accumulation near the resonance that is consistent with the prominent phase branch cut observed on the $k_x > 0$ side. In other words, the branch cut that appears only for $k_x > 0$ in Figure 2 originates from a topological rearrangement of the zero-pole pair, triggered by the zero point crossing the real-frequency axis.

To further examine the evolution of the zero point with $k_x$, Figure 4(d) plots the Im($f$) of the zero point as a function of $k_x$ for different values of the structural asymmetry $\delta d$. As $k_x$ approaches the Γ point, Im($f$) of the zero point converges toward the origin, indicating the appearance of a transmission zero at the Γ point. Figure 4(e) shows the magnitude of the Lorentz-oscillator coupling strength $|B|$ as a function of $k_x$. Over the range examined, $|B|$ increases monotonically with $k_x$: for $k_x < 0$, $|B|$ remains small and the qBIC is only weakly coupled to the radiation channel, corresponding to the zero point not yet having crossed the real-frequency axis, and thus the phase does not exhibit a full 2π winding. When $k_x$ crosses the Γ point into the $k_x > 0$ region, $|B|$ increases significantly; the enhanced coupling "pulls" the zero point to the opposite side of the real-frequency axis, thereby causing the real-frequency sweep to cross the branch cut and inducing a strong phase variation. The phase of $B$ varies slowly versus $k_x$ with no abrupt change near the Γ point as shown in Figure 4(f). This behavior demonstrates that the observed phase singularity and abrupt phase evolution in the transmission response are not caused by a rapid variation in the coupling phase arg($B$), but instead originate from the strong $k_x$-dependent modulation of the coupling magnitude $|B|$ between the obliquely incident wave and the qBIC resonance.

On the other hand, the magnitude and phase of the background term $A$ remain nearly invariant for $k_x \in [-0.05, 0.05]$ $(2\pi/a)$ as shown in Figure S9 in Supplementary Materials. This contrast clearly indicates that both the migration of the zero point in the complex-frequency plane and the crossing of the branch cut through the real axis of frequency exclusively for $k_x > 0$ originate from the strongly directional coupling with the resonance, rather than from changes in the background scattering. It should be noted that the zero-pole distribution and its evolution depend on the symmetry-breaking direction. If the lower air slots shift toward the $-x$ direction instead of the $+x$ direction, the zero-and-pole pair will cross the real axis of frequency in the region $k_x < 0$ correspondingly, and the spiral phase profile will also be reversed, as shown in Figure S10 of the Supplementary Materials.

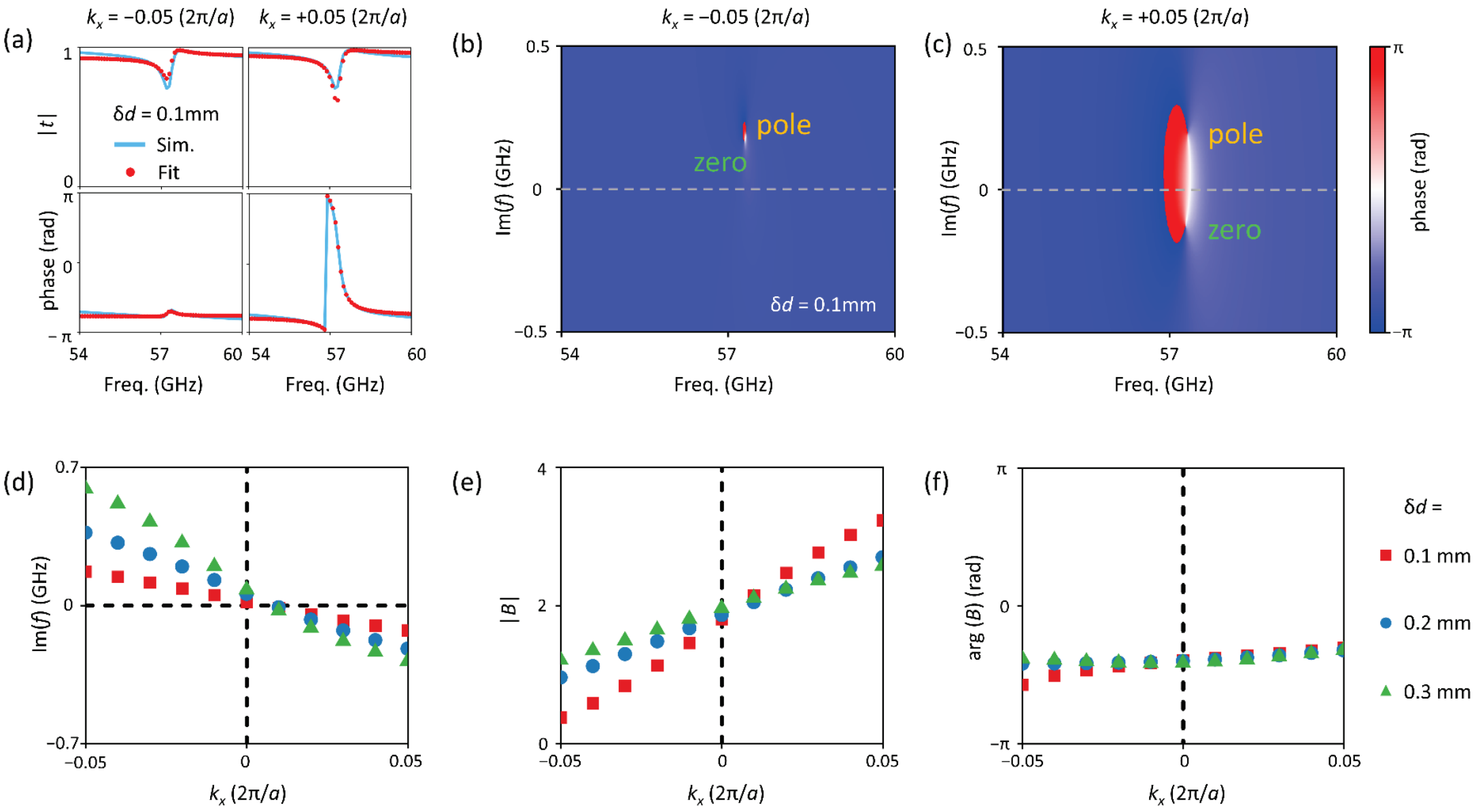

**Figure 4. Scattering characteristics and parameter evolution based on the Lorentz oscillator model.** (a) Amplitude and phase of the transmission coefficient at $k_x = -0.05$ $(2\pi/a)$ and $+0.05$ $(2\pi/a)$ for $\delta d = 0.1$ mm, comparing numerical simulations (blue solid lines) with the Lorentz-model fitting (red dots). (b) - (c) Transmission-phase distributions in the (Re($f$), Im($f$)) space for $k_x = -0.05$ $(2\pi/a)$ and $+0.05$ $(2\pi/a)$. The locations of the zero and pole points are marked. (d) Evolution of Im($f$) of the zero point as a function of $k_x$. (e) - (f) Variation of magnitude and phase of $B$ of the Lorentz-resonant term versus $k_x$. Red squares, blue circles, and green triangles correspond to $\delta d = 0.1, 0.2, 0.3$ mm, respectively.

**E. Experimental verification and material-loss effects**

To verify the theoretical analyses and numerical simulations presented above, we constructed the millimeter-wave free-space transmission measurement system shown in Figure 5(a) to accurately measure the transmission amplitude and phase of the sample. The experiment employs a millimeter-wave vector-network-analysis system as the test platform, consisting of a vector network analyzer (VNA), a pair of millimeter-wave frequency-extension modules (Freq. Ext. Mod.) operating in the 50 - 75 GHz band, a pair of linearly polarized antennas operating in the corresponding frequency range, and a rotation stage used to mount the sample and vary the angle of incidence. Figure 5(b) shows the overall and detailed views of the sample in both side and top views. The photograph of the experimental setup is provided in Supplementary Materials Figure S11.

Figures 5(c) and 5(d) show distributions of transmission amplitude and phase in $\theta$ - $f$ space obtained from numerical simulations with a material loss tangent of tan$\delta$ = 0.006. The reconstructed amplitude and phase of the STOV are shown in the Supplementary Materials Figure S12. Figures 5(e) and 5(f) present the corresponding experimental results. The resonance frequency, angular dispersion trajectory, and the sharp phase jump near the resonance exhibit strong agreement between simulation and experiment. In Figures 5(c) and (e), the regions where the transmission amplitude approaches zero are marked with white dashed loop. As demonstrated in Figure S13 of the Supplementary Materials, for a reasonable range of material loss tangents, the existence of the transmission zero is robust against moderate material loss.

Notably, the measured near-zero transmission minimum and associated phase singularity occur at a non-zero angle of incidence. In an ideal scenario without any material loss as considered in Figure 2, the transmission zero is pinned at the Γ point because, under normal incidence, the background transmission can be exactly canceled by the forward scattering mainly contributed by the magnetic dipole moment $MD_y$, guaranteed by the $C_2$ rotational symmetry in the $x$ - $z$ plane. Once the material loss is introduced, however, the forward scattering strength is reduced, so that the destructive interference at normal incidence becomes incomplete and the transmission dip can no longer reach zero at $k_x = 0$. To recover a complete cancellation, the forward scattering needs to be enhanced, so that an oblique incidence is required. As discussed in Figure 3, for an oblique incidence, or a non-zero $k_x$, the excited electric dipole component $ED_x$ interferes with $MD_y$ in the far field, enhancing the forward scattering and suppressing the backward scattering when $k_x > 0$. This enhanced forward radiation enables the background transmission to be fully canceled again at a certain angle of incidence. As a result, with an increasing loss tangent, the transmission zero detaches from the Γ point more and more, and the associated phase-singularity moves away from $k_x = 0$, consistent with the zero-pole evolution shown in Supplementary Materials Figure S14.

We also note that the experimentally observed transmission-zero frequency is slightly higher than that predicted by the numerical simulation. This deviation is mainly attributed to practical imperfections, including uncertainties in the material refractive index, fabrication tolerances in the slot depth, finite-size effects, and structural nonuniformity. As shown by the error analysis in the Supplementary Materials Figure S15, the transmission-zero frequency is sensitive to variations in the refractive index and slot-depth-related parameters. These imperfections may also partially lift the transmission zero and smear the associated phase singularity, resulting in a less pronounced phase jump in the experiment than in the simulation. Nevertheless, the transmission zero and the corresponding phase-singularity behavior remain robust within a reasonable range of material loss and structural nonuniformity, confirming that the observed STOV response is an intrinsic and robust feature of the designed structure rather than an accidental consequence of idealized parameters. The weak stripe-like features observed in the experimental amplitude map are mainly caused by residual multiple reflections in the free-space measurement setup, especially among the horn antennas, the sample, and nearby components, which give rise to unavoidable standing-wave-like interference in the measured spectra. The momentum-frequency signatures are also confirmed by another sample with an increased $\delta d$ = 0.15 mm, in which a transmission minimum of approximately 0.015 and a robust branch-cut feature are experimentally observed, as shown in Figure S16 of Supplementary Materials.

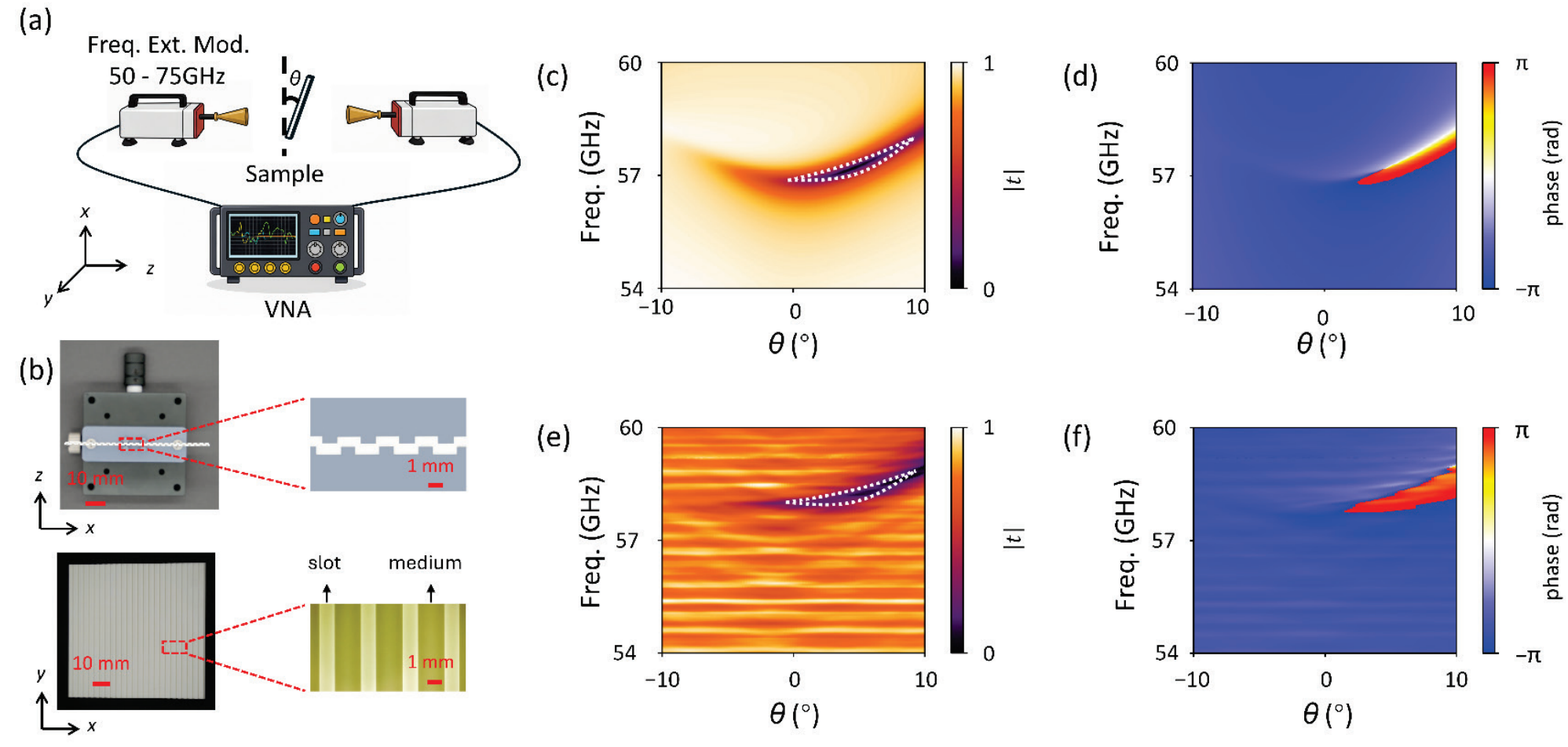

**Figure 5. Experimental setup and comparison of simulated and measured transmission characteristics in the $\theta$ - $f$ space.** (a) The 50 - 75 GHz frequency-extension modules with horn antennas connected to the VNA to measure the S parameters in the corresponding frequency range. $\theta$ is the oblique incidence angle. (b) Side view of the sample and its microstructure (top); top view of the sample and its microstructure (bottom). (c) - (d) Simulated distributions of the transmission amplitude and phase in the $\theta$ - $f$ space, with the material loss tangent set to 0.006. (e) - (f) Corresponding experimentally measured transmission amplitude and phase distributions in the $\theta$ - $f$ space. The regions where the transmission amplitude approaches zero are marked with white dashed loop. The measured minimum $|t|$ in the map of (e), inside the white dashed loop, is 0.02.

## III. CONCLUSION

In this work, we propose and investigate a qBIC-based mechanism for generating spatiotemporal optical vortices (STOVs) in all-dielectric bilayer metagratings. By introducing a tunable lateral shift between the two layers, a symmetry-protected BIC is converted into a radiative qBIC, enabling robust vortex-like and spiral-phase waveforms in the $x$ - $T$ domain under excitation by a spatiotemporal Gaussian pulse. Through multipole analysis, we show that slight symmetry breaking activates the orthogonal electric and magnetic dipoles ($ED_x$ and $MD_y$); their Kerker-type interference redistributes the radiated power between the two half-spaces and leads to strongly asymmetric downward and upward radiation. To elucidate the phase-singularity mechanism underlying STOV formation, we further employ an effective Lorentz-oscillator description in which a narrow qBIC resonance coherently interferes with a smooth background channel. This model reproduces the transmission spectra and reveals a clear physical origin of the phase singularity: as the coupling strength $|B|$ increases for oblique momentum, the transmission zero migrates in the complex-frequency plane and crosses the real-frequency axis, thereby producing an almost $2\pi$ phase winding and the prominent phase branch cut observed in the momentum-frequency response. Finally, free-space measurements in the $\theta$ - $f$ space reproduce the key dispersion and phase signatures, confirming the transmission zero and the branch-cut behavior. Although experimentally demonstrated at millimeter-wave frequencies, the all-dielectric mechanism is readily scalable to higher frequency regimes and compatible with quantum emitters. Notably, the interlayer lateral shift serves as a simple knob to program a family of tailored wave packets, offering a potentially reconfigurable platform in which both the spatial and temporal waveforms can be tuned, motivating future explorations.

## SUPPLEMENTARY MATERIAL

See [Supplementary Material] for the discussion on qBIC Q-factor scaling, transmitted magnetic-field distributions, $k_x$ - $f$ transmission evolution with structural asymmetry, negative-asymmetry STOV generation, second-order STOVs from cascaded metagratings, slot-depth-mismatch-induced radiation imbalance, far-field radiation patterns, tunable directional radiation, background-term magnitude and phase, zero-pole rearrangement under reversed symmetry breaking, free-space S-parameter measurement setup, lossy oblique-incidence spatiotemporal response, loss-dependent transmission spectra and zero trajectories, effects of material and structural variations, and enhanced experimental observation of the branch-cut response.

## ACKNOWLEDGMENTS

This research was supported by the National Natural Science Foundation of China (No. 12304348, 62505236), Guangdong Basic and Applied Basic Research

Foundation (No. 2025A1515011470), Guangdong University Featured Innovation Program Project (2024KTSCX036), Guangdong Provincial Project (2023QN10X059), Guangzhou-HKUST(GZ) Joint Funding Program (2025A03J3783), HKUST-HKUST(GZ) 20 for 20 Cross-campus Collaborative Research Scheme (G011), Guangzhou Municipal Science and Technology Project (2024A04J4351), and the Fundamental Research Funds for the Central Universities (WUT:104972024RSCbs0023). The authors would like to acknowledge the Wave Functional Metamaterial Research Facility (WFMRF) of The Hong Kong University of Science and Technology (Guangzhou) for the experimental support.

## AUTHOR DECLARATIONS

The authors have no conflicts to disclose.

**Supplementary Materials on**

# Spatiotemporal Optical Vortices From All-Dielectric Bilayer Metagratings

Ken Qin[1], Shijie Kang[1], Aoning Luo[1], Yiyi Yao[1], Xiexuan Zhang[1], Hanchuan Chen[1], Yahan Xiao[2], Yangsong Ye[3], Junqing Shi[1], Xusheng Xia[4,*], Haitao Li[1,*], and Xiaoxiao Wu[1,*]

[1]*Modern Matter Laboratory and Advanced Materials Thrust, The Hong Kong University of Science and Technology (Guangzhou), Nansha, Guangzhou 511400, Guangdong, China*

[2]*School of Integrated Circuit Science and Engineering, University of Electronic Science and Technology of China, Chengdu 610054, China*

[3]*Center for Information Photonics and Energy Materials, Shenzhen Institute of Advanced Technology, Chinese Academy of Sciences, Shenzhen 518055, China*

[4]*School of Physics and Mechanics, Wuhan University of Technology, Wuhan 430070, China*

[*]*To whom correspondence should be addressed. E-mails: xsxia@whut.edu.cn (X. Xia), haitaoli@hkust-gz.edu.cn (H. Li) and xiaoxiaowu@hkust-gz.edu.cn (X. Wu)*

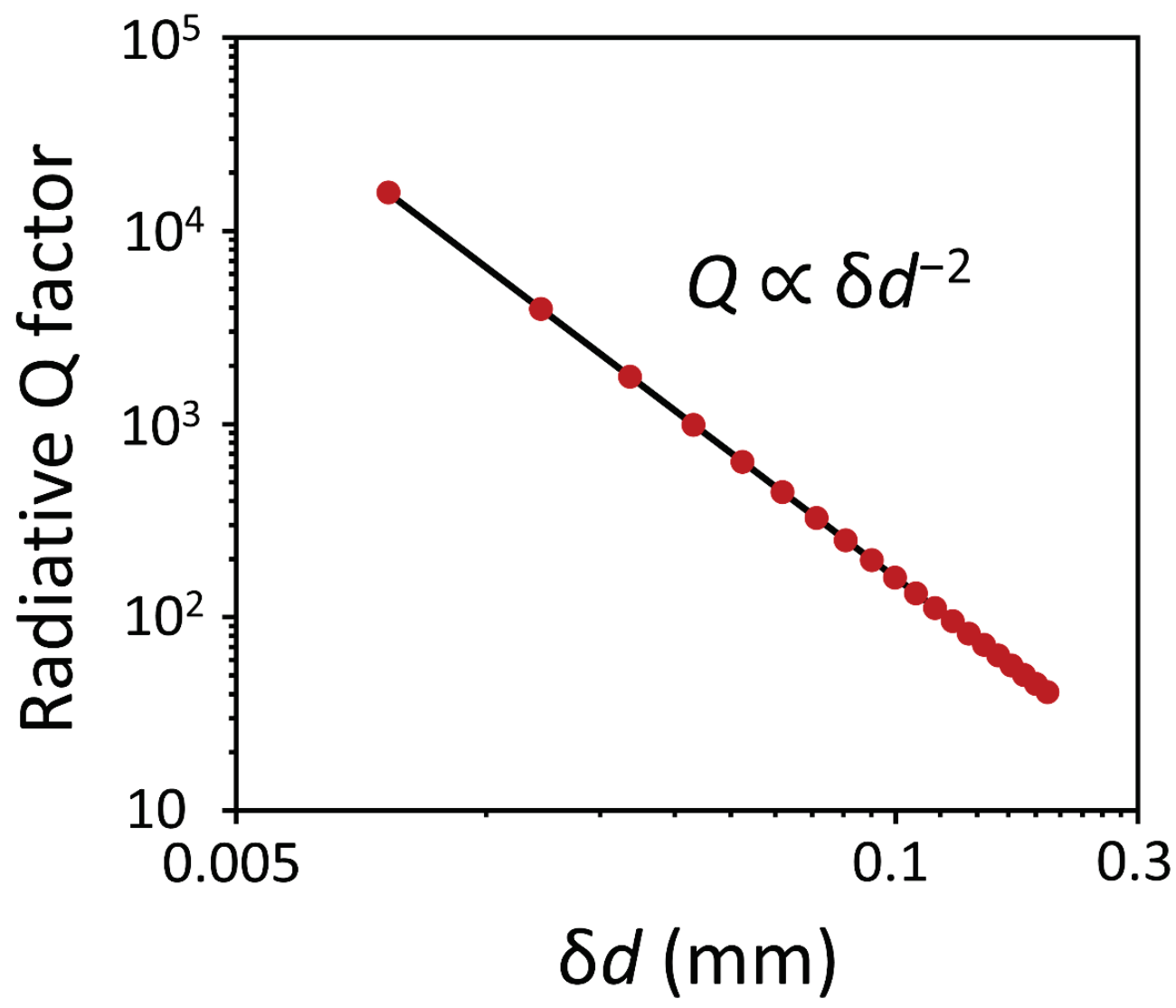

**Figure S1. Radiative Q-factor of the qBIC as a function of the asymmetry parameter δ*d* plotted in log-log scale.** Red dots represent the numerically calculated results, while the black solid line denotes a power-law fitting. The radiative Q-factor follows an inverse quadratic scaling, $Q \propto \delta d^{-2}$, which is a hallmark of SP-BICs (with topological charge 1) perturbed by structural asymmetry. This behavior is fully consistent with previous theoretical and numerical studies on asymmetric metasurfaces and photonic structures, confirming the qBIC nature of the observed resonance.[1]

Figure S1: Log-log plot of the radiative quality factor versus lateral asymmetry shows an inverse quadratic scaling. The numerical data and fitted curve confirm that the resonance behaves as a qBIC induced by structural asymmetry.

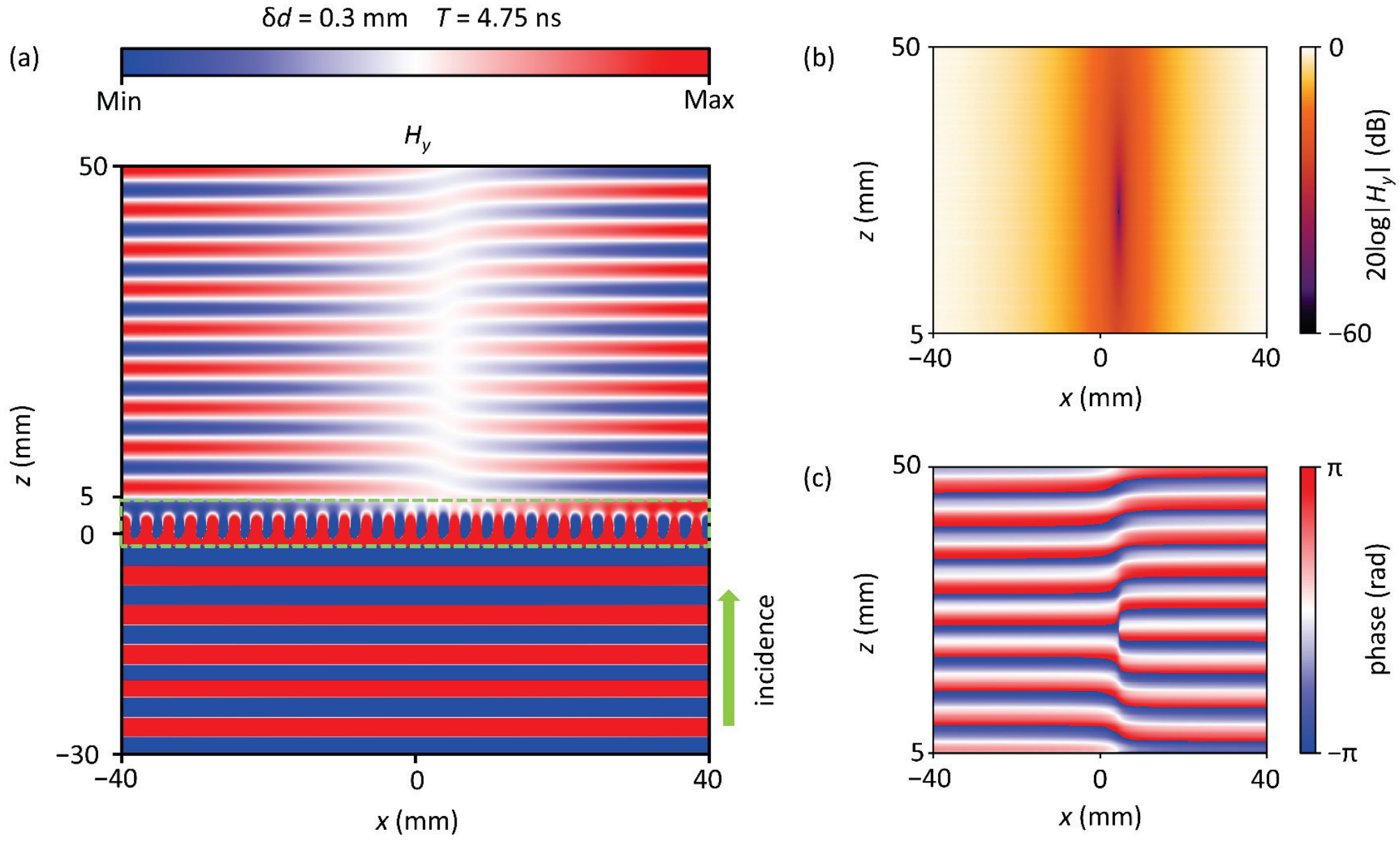

**Figure S2. Magnetic-field distributions for δ*d* = 0.3 mm at a fixed time *T* = 4.75 ns.** (a) Spatial distribution of $H_y$ in the *x*-*z* plane. The green dashed box indicates the structural region, and the green arrow marks the direction of incidence along the positive *z* axis. Due to the presence of a phase singularity, the field exhibits a characteristic fork-like pattern in the *x*-*z* plane. (b) Magnitude distribution of the reconstructed field obtained via the Hilbert transform.[2,3] (c) Corresponding phase distribution. The fork-shaped phase pattern reveals a phase singularity, around which the accumulated phase winding is $-2\pi$, indicating a winding number of $-1$.

Figure S2: Magnetic-field maps show the transmitted field for an asymmetric bilayer metagrating at a fixed time. The reconstructed magnitude and phase distributions reveal a fork-like pattern and a phase singularity with a winding number of $-1$.

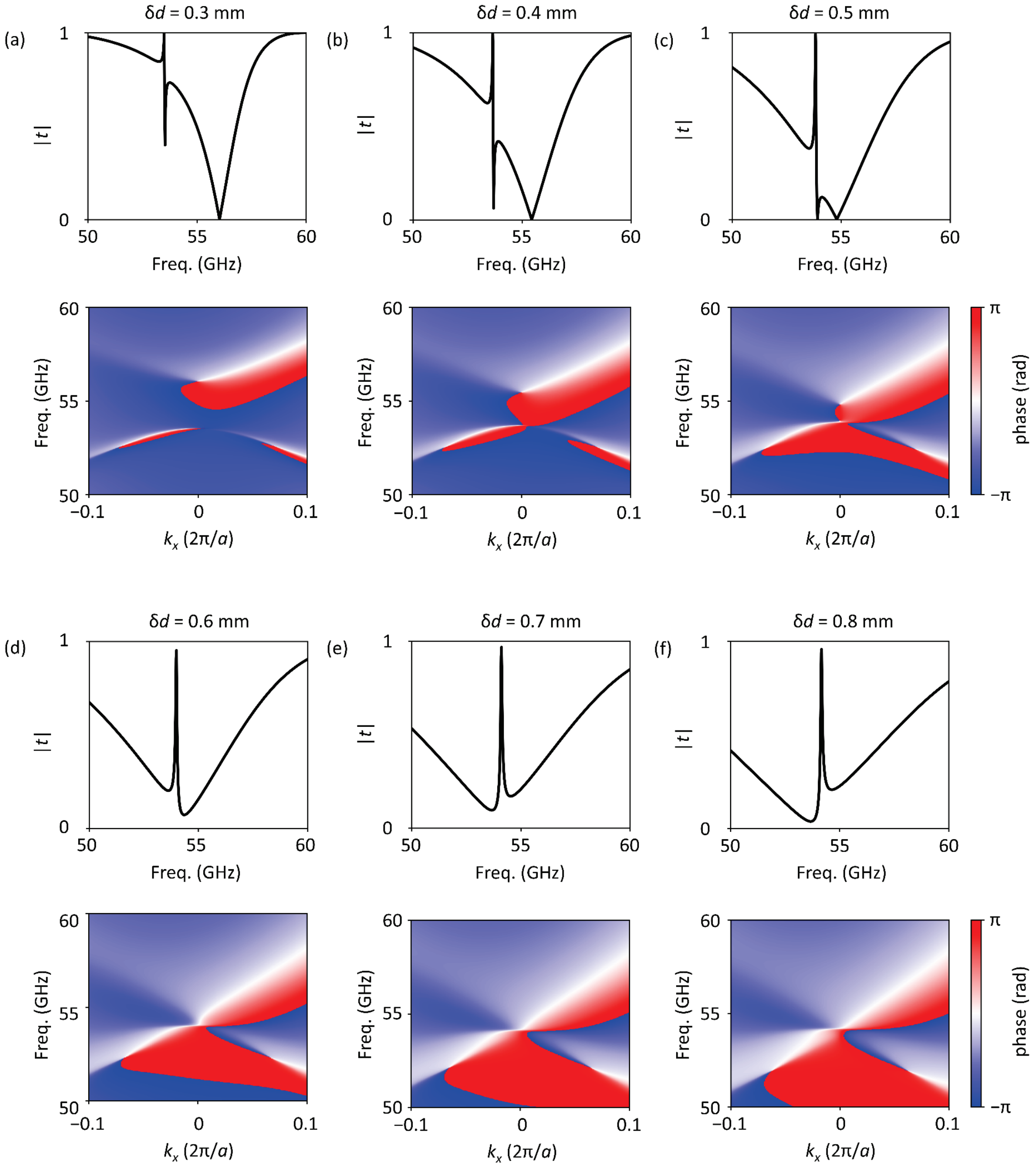

**Figure S3. Evolution of the transmission response in the $k_x$ - $f$ space for different asymmetry parameters δ*d*.** Panels (a) - (f) correspond to δ*d* = 0.3, 0.4, 0.5, 0.6, 0.7 and 0.8 mm, respectively. The upper row in each panel shows the absolute value of the transmission coefficient $|t|$ as a function of frequency at normal incidence, while the lower row presents the corresponding transmission phase distribution in the $k_x$ - $f$ space. For small δ*d*, two clearly separated resonant modes give rise to distinct phase singularities and associated phase vortices in the $k_x$ - $f$ plane. As δ*d* increases, the two modes become progressively closer, leading to mode hybridization and the merging and annihilation of the originally independent

phase singularities. Consequently, the phase vortices disappear, and the phase distribution evolves into a smooth, continuous profile.

Figure S3: Transmission spectra and phase maps show the evolution of the momentum-frequency response as the asymmetry parameter increases. The initially separated phase singularities merge and disappear because of mode hybridization at larger structural offsets.

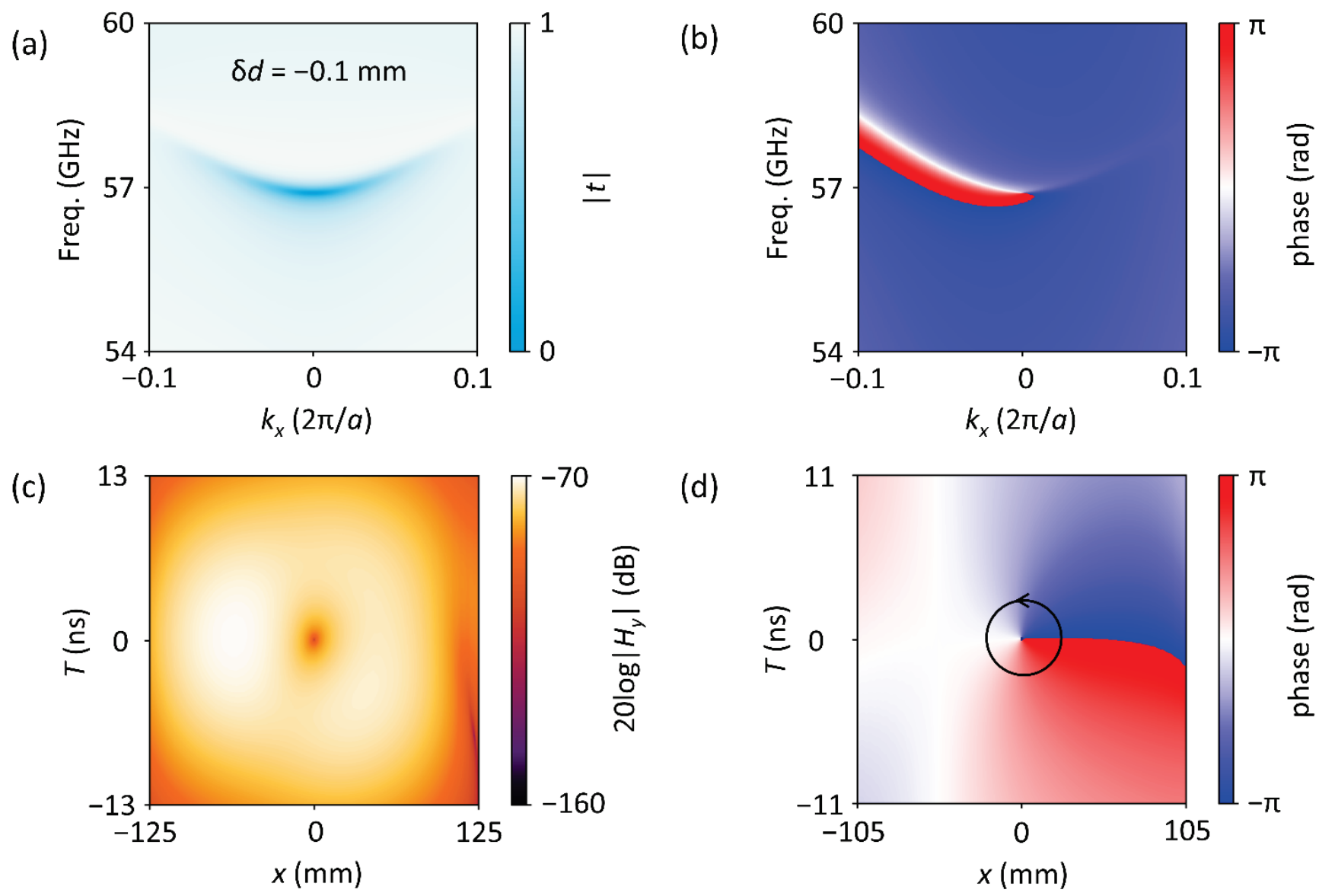

**Figure S4. Momentum-frequency-domain transmission response and space-time-domain field characteristics for a negative asymmetry parameter δ*d* = −0.1 mm.** (a) The distribution of the amplitude of the transmission coefficient $|t|$ in the $k_x$ - $f$ space, showing that the transmission dip (zero) persists despite the reversed asymmetry. (b) Corresponding transmission phase map, where a phase singularity is observed near the resonance. In contrast to the positive-$\delta d$ case, the phase winding direction is reversed, yielding a total phase accumulation of $2\pi$ along a counterclockwise loop and thus indicating a winding number of +1. (c) Magnitude distribution of the magnetic field component in the $x$-$T$ space. (d) Corresponding phase distribution of the magnetic field component; the circle marks the vortex core and the arrow denotes the reversed winding direction. The accumulated phase along a counterclockwise loop of $2\pi$ confirms a winding number of +1. This behavior indicates that while the existence of the vortex is governed by the survival of phase singularities, the sign of the winding number is controlled by the direction of symmetry breaking, providing an additional degree of robustness and tunability in the STOV generation.

Figure S4: Momentum-frequency and space-time-domain maps show the response for a negative asymmetry parameter. The transmission zero remains present, but the phase winding direction reverses, producing a spatiotemporal optical vortex with winding number +1.

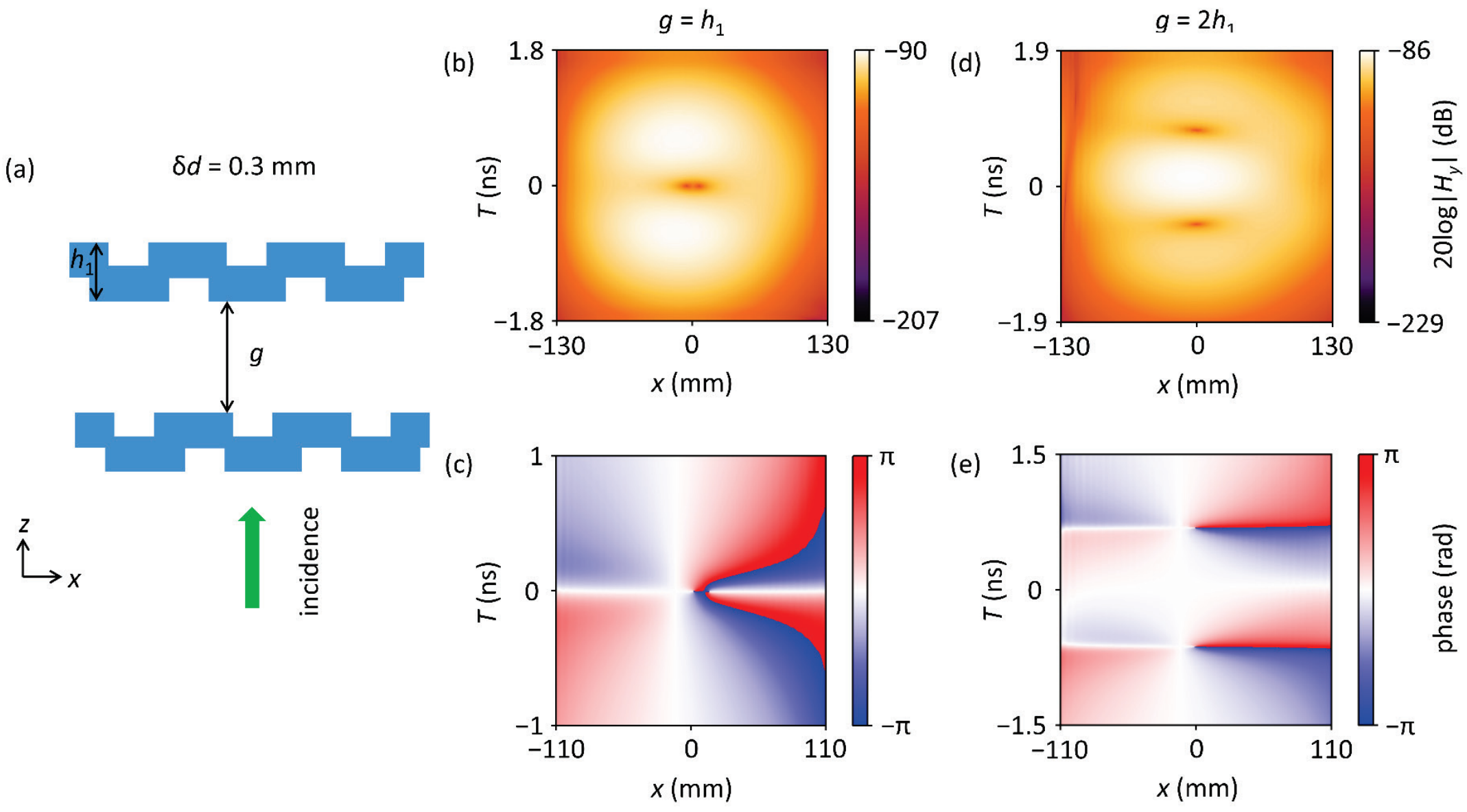

**Figure S5. FDTD demonstration of second-order STOV generation using cascaded bilayer metagratings.** (a) Schematic of two identical bilayer metagratings stacked along the propagation direction with an air gap $g$, where $h_1$ denotes the thickness of a single bilayer metagrating. The incident wave propagates along the $+z$ direction. (b) and (c) Magnitude and phase distributions of the transmitted $H_y$ field in the $x$ - $T$ domain for $g = h_1$. (d) and (e) Corresponding magnitude and phase distributions for $g = 2h_1$. In both cases, the transmitted field shows the characteristic features of a second-order STOV, demonstrating the feasibility of generating higher-order STOVs by cascading multiple bilayer metagrating units. The incident wave packet parameters are $w_x = 45$ mm, $\tau = 1$ ns and $\delta d = 0.3$ mm.

Figure S5: Simulated cascaded bilayer metagratings generate second-order spatiotemporal optical vortices. Magnitude and phase maps in the space-time domain show higher-order vortex features for two different air-gap distances between the bilayer metagratings.

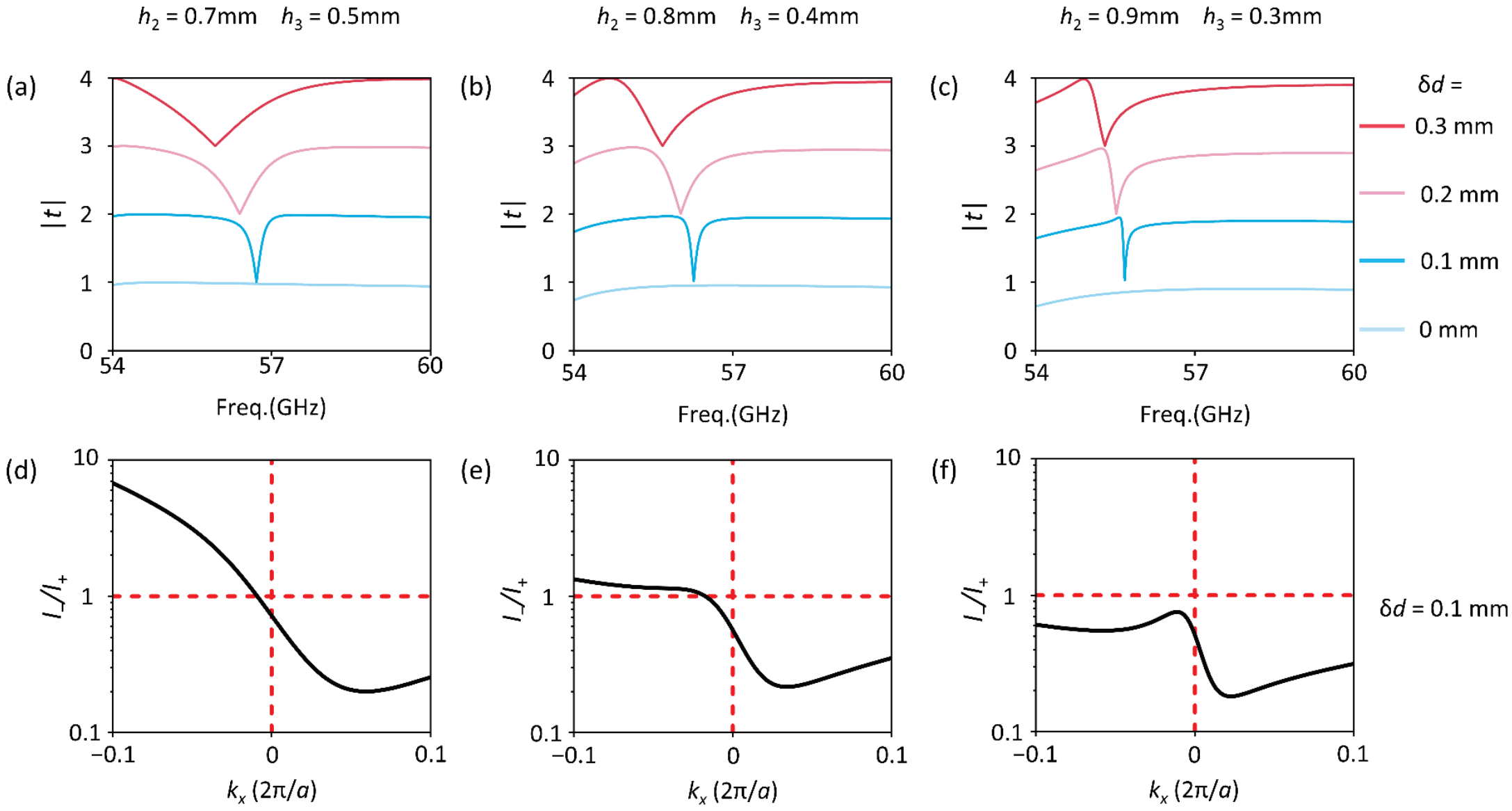

**Figure S6. Unequal downward and upward radiation induced by slot-depth mismatch as a function of $k_x$.** (a) - (c) Simulated transmission magnitude $|t|$ for three representative geometries with ($h_2$, $h_3$) = (0.7, 0.5) mm, (0.8, 0.4) mm and (0.9, 0.3) mm, respectively. Curves are color-coded for different asymmetry parameters $\delta d$ (0 - 0.3 mm). (d) - (f) Poynting-vector-flux ratio $I_-/I_+$ as a function of $k_x$ for $\delta d$ = 0.1 mm for the same three geometries. The red dashed lines mark $k_x = 0$ and $I_-/I_+ = 1$. Unlike the case that $h_2 = h_3$, here, the upward and downward radiation are unequal at $k_x = 0$.

Figure S6: Simulated transmission spectra and Poynting-vector-flux ratios show the effect of unequal upper and lower slot depths. Slot-depth mismatch breaks the balance between downward and upward radiation at the Γ point.

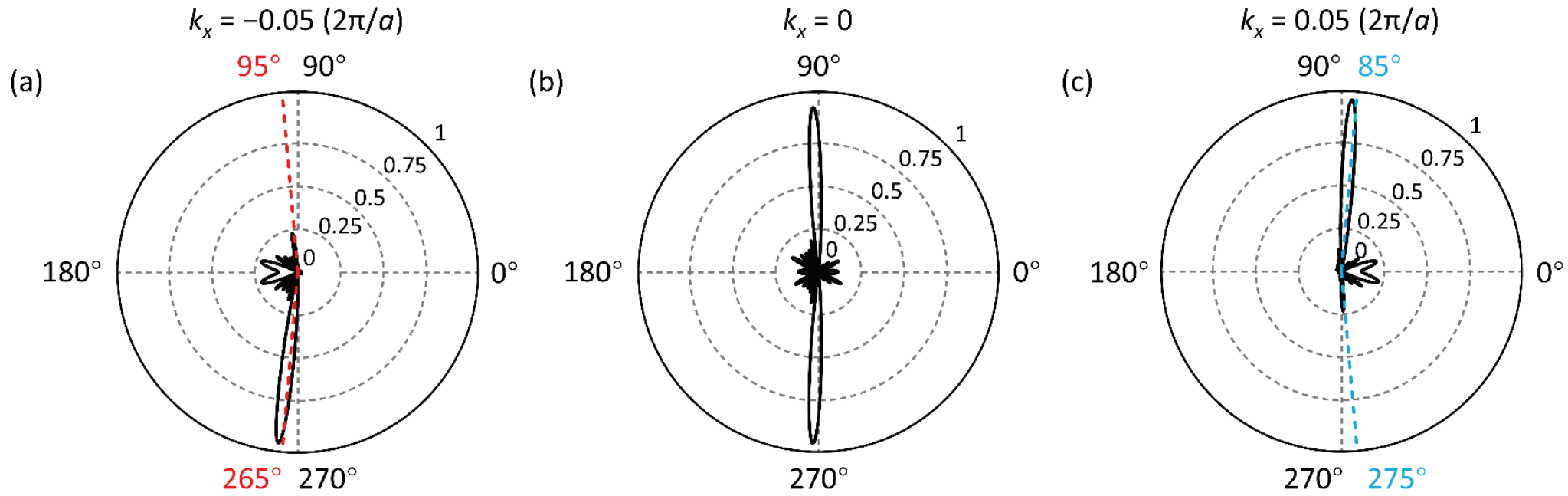

**Figure S7. Normalized far-field radiation patterns of the qBIC mode.** Individually normalized angular distributions of the far-field radiation power in the *x*-*z* plane are calculated from 11 unit cells for $\delta d$ = 0.1 mm at three representative transverse wavevectors: (a) $k_x$ = −0.05 (2π/*a*) (red dashed line), (b) $k_x$ = 0, and (c) $k_x$ = +0.05 (2π/*a*) (blue dashed line). At $k_x$ = 0, the pattern is symmetric with the main lobe along 90° and 270°. For $k_x$ = −0.05 (2π/*a*), the dominant downward lobe tilts to 265°, accompanied by a weaker upper-side lobe around 95°. Conversely, for $k_x$ = +0.05 (2π/*a*), the dominant upward lobe tilts to 85°, with a corresponding weaker lower-side lobe around 275°. These $k_x$-dependent angular shifts reflect the breaking of mirror symmetry at off-Γ points and visualize the Kerker-type multipolar interference responsible for directional emission.

Figure S7: Normalized far-field radiation from 11 unit cells is symmetric at the Γ point, with the main emission directed vertically. Away from the Γ point, the main lobe tilts upward or downward with a weaker opposite side lobe, revealing directional radiation caused by broken mirror symmetry and multipolar interference.

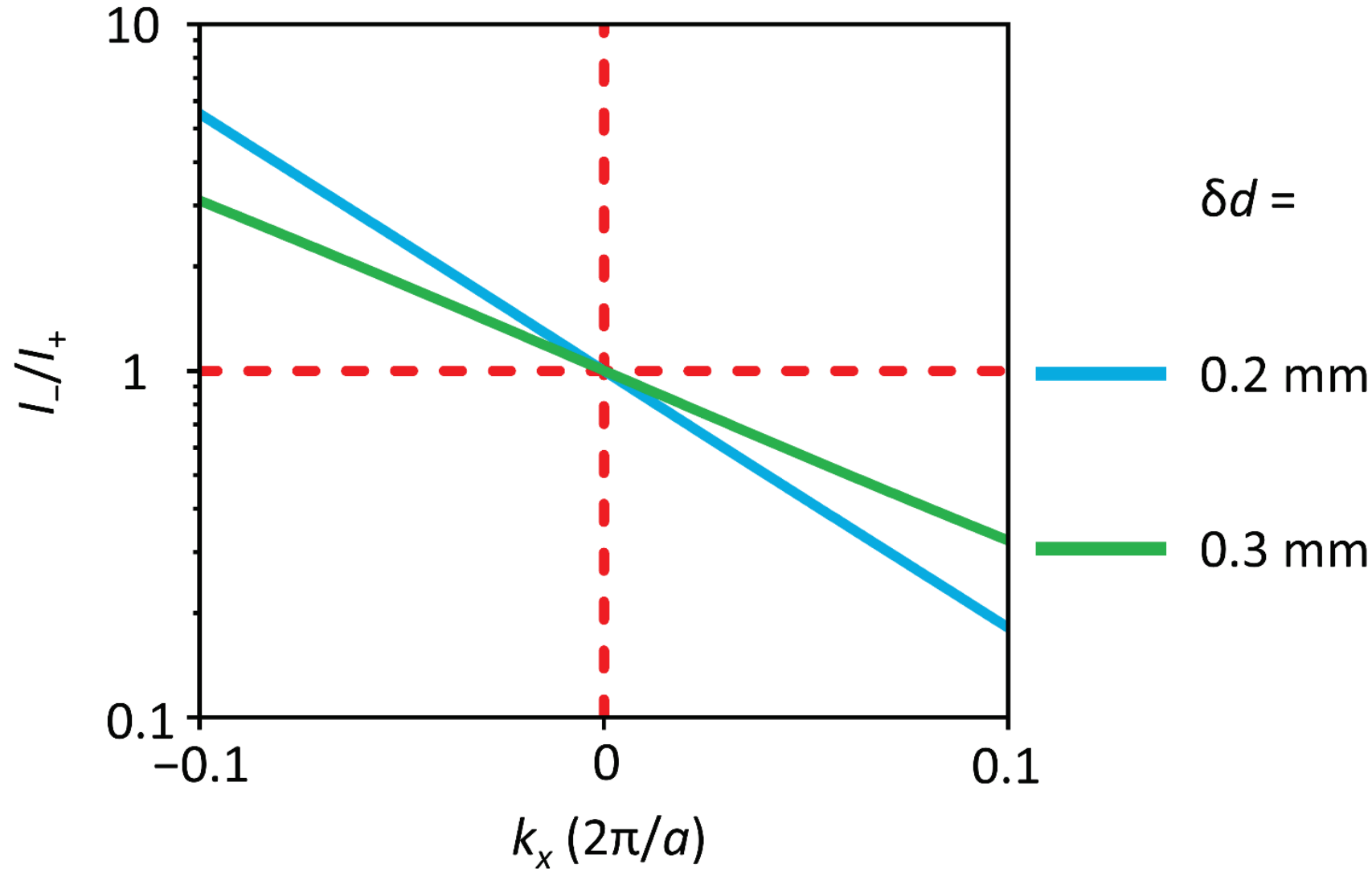

**Figure S8. Dependence of the radiation asymmetry slope on the structural asymmetry parameter δ*d*.** Ratio of Poynting-vector-flux $I_{–}/I_{+}$ plotted as a function of $k_x$ for different values of δ$d$. The logarithmic representation shows that increasing δ$d$ results in a flatter slope of log($I_{–}/I_{+}$), indicating a reduced exponential sensitivity of the radiation asymmetry to $k_x$. These results confirm that the degree of directional radiation discussed in Figure 3(f) can be systematically tuned by controlled symmetry breaking.

Figure S8: Poynting-vector-flux ratios plotted against Bloch wavevector show how directional radiation changes with the structural asymmetry parameter. Larger asymmetry produces a flatter logarithmic slope, indicating reduced sensitivity of radiation imbalance to momentum.

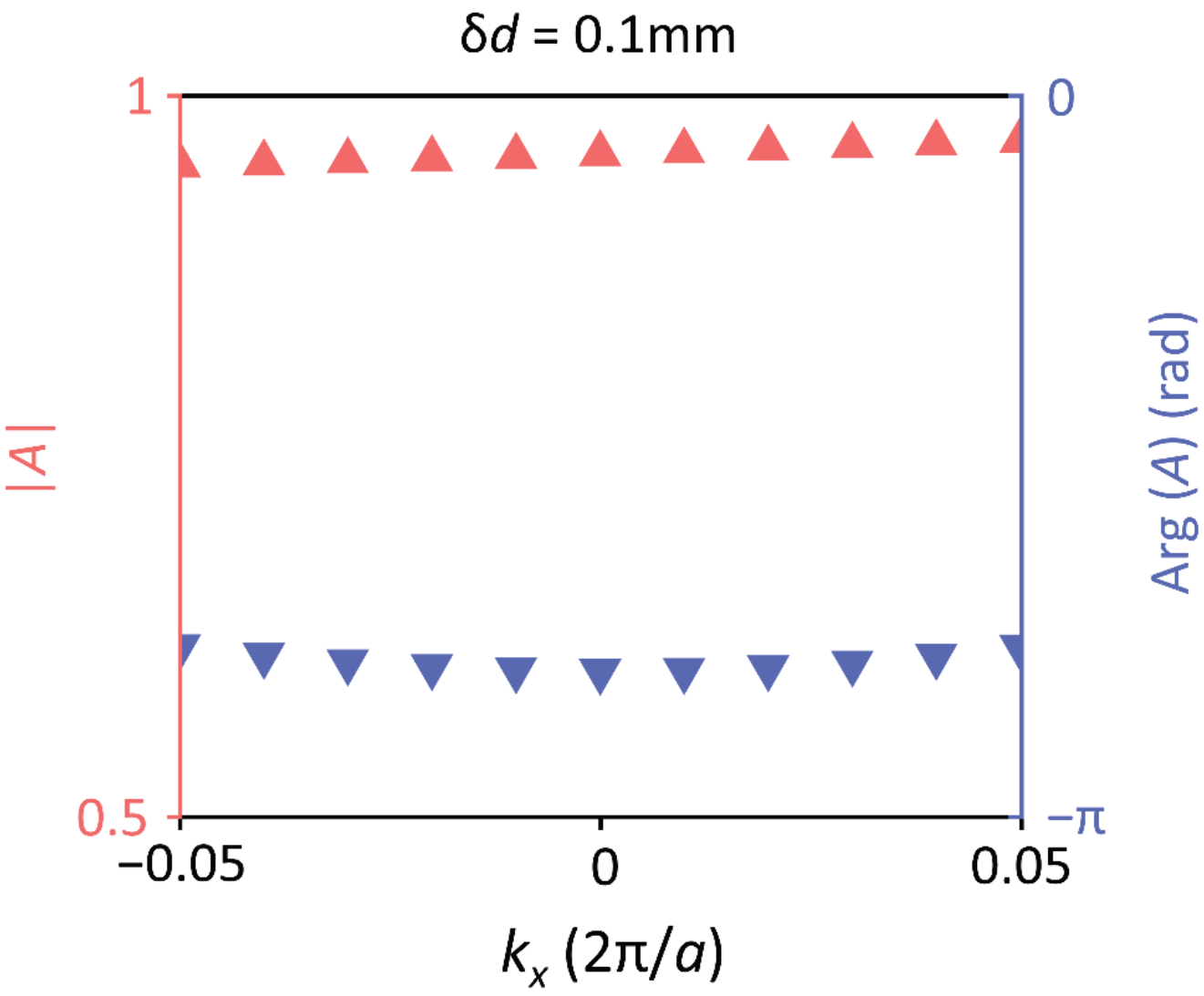

**Figure S9. Magnitude (orange upright triangles) and phase (purple inverted triangles) of background term $A$ as functions of $k_x$ when $\delta d$ = 0.1 mm.** Both magnitude and phase of $A$ vary weakly with $k_x$ over the examined range, indicating that the background contribution remains nearly constant and does not play a dominant role in the observed transmission phase evolution.

Figure S9: The magnitude and phase of the background scattering term vary weakly with Bloch wavevector. This nearly constant response indicates that background scattering is not the main cause of the observed transmission phase evolution.

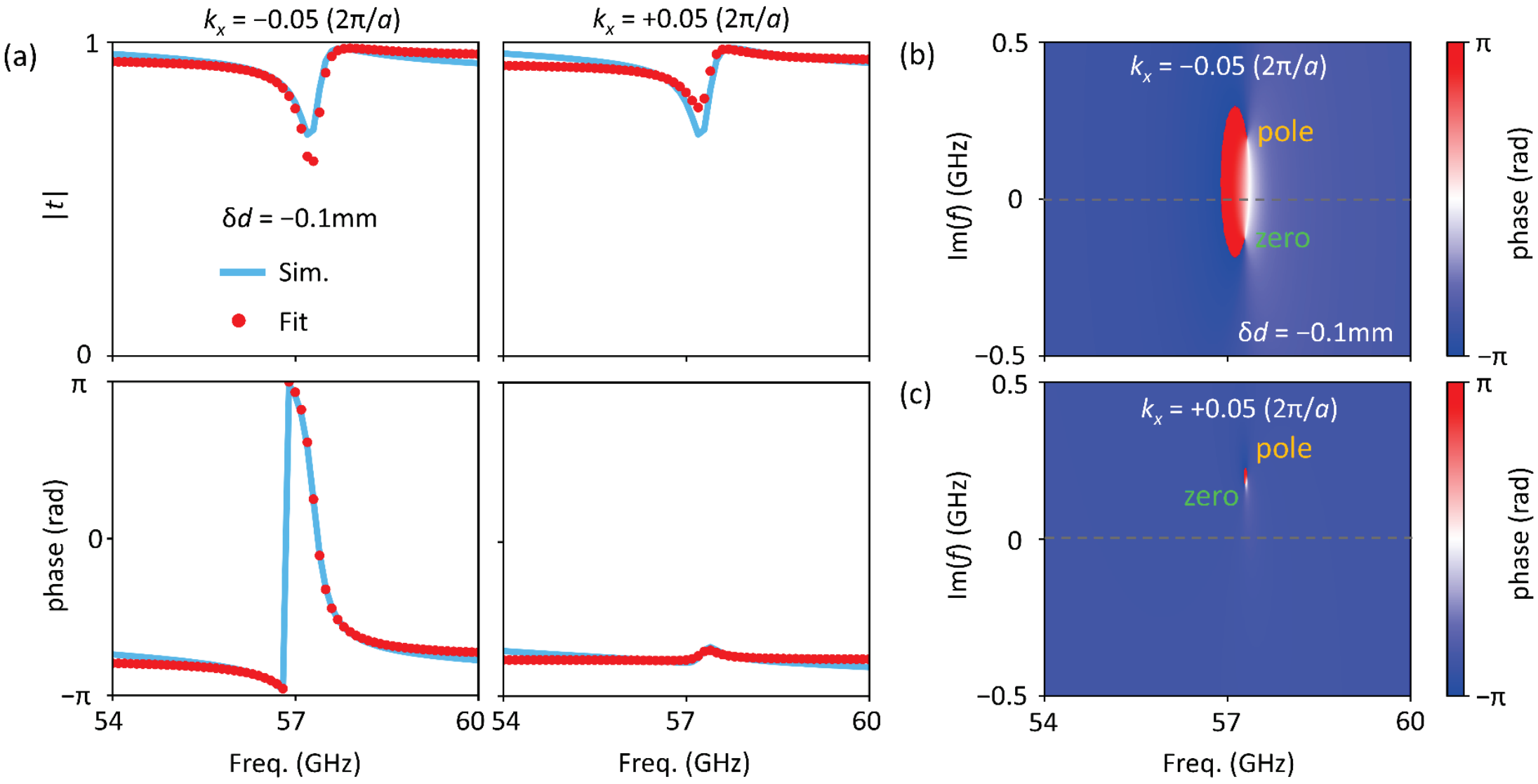

**Figure S10. Zero-pole rearrangement and phase singularities under the reversed symmetry-breaking direction.** (a) Transmission amplitude $|t|$ (top) and phase (bottom) at $k_x = -0.05\ (2\pi/a)$ and $k_x = +0.05\ (2\pi/a)$ for $\delta d = -0.1$ mm, comparing full-wave simulations (blue solid lines) with Lorentz-oscillator-model fitting (red dots). (b) - (c) Phase maps of the analytically continued transmission coefficient in the complex-frequency plane (Re($f$), Im($f$)) for $k_x = -0.05\ (2\pi/a)$ and $k_x = +0.05\ (2\pi/a)$, respectively. In this reversed symmetry-breaking configuration, the topological behavior is swapped compared with the main-text case: for $k_x < 0$, the zero point migrates across the real-frequency axis into the opposite half-plane while the pole point remains essentially fixed, so that the sweep along the real-frequency axis effectively passes between the zero-and-pole branch cut. In contrast, for $k_x > 0$ the zero and pole points lie on the same side of the real-frequency axis, the real-axis sweep does not cross between them, and the transmission phase exhibits only a small jump without a phase wrapping.

Figure S10: Lorentz-model fitting and complex-frequency phase maps show the zero-pole behavior for the reversed symmetry-breaking direction. The topological phase response shifts to the opposite momentum side compared with the main-text configuration.

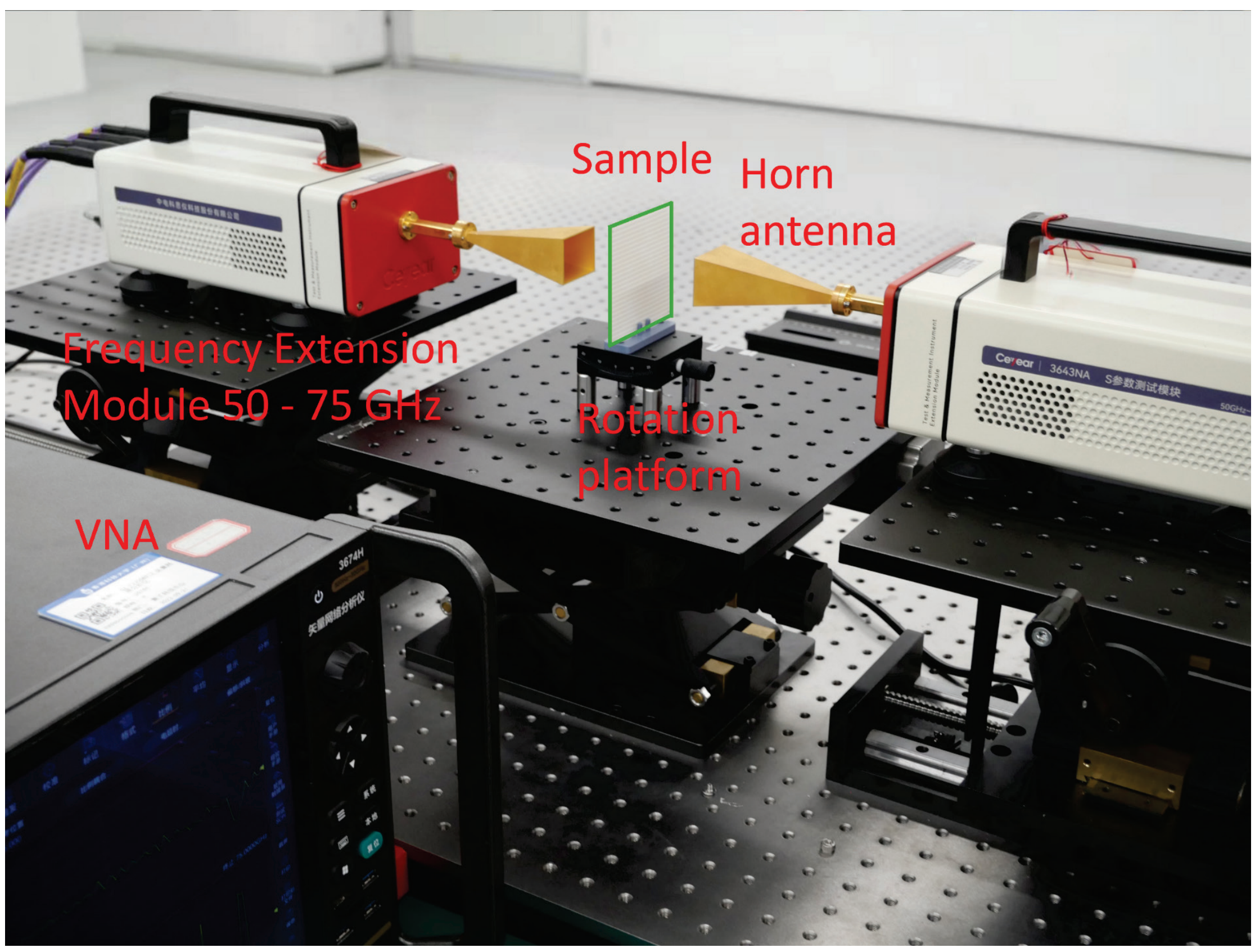

**Figure S11. Experimental free-space S-parameter measurement setup.** A VNA equipped with a pair of frequency-extension modules (50 - 75 GHz, Ceyear, China) provides the free-space transmission measurements. The radiated signal is launched and collected by a pair of horn antennas, with the metagrating sample mounted between them. The sample is fixed on a motorized rotation (goniometric) platform, which allows tuning the incidence direction (pitch angle) to measure S-parameters under different oblique incidences.

Figure S11: Photograph of the free-space millimeter-wave measurement setup shows the vector network analyzer, frequency-extension modules, horn antennas, metagrating sample, and motorized rotation platform used to measure angle-dependent transmission.

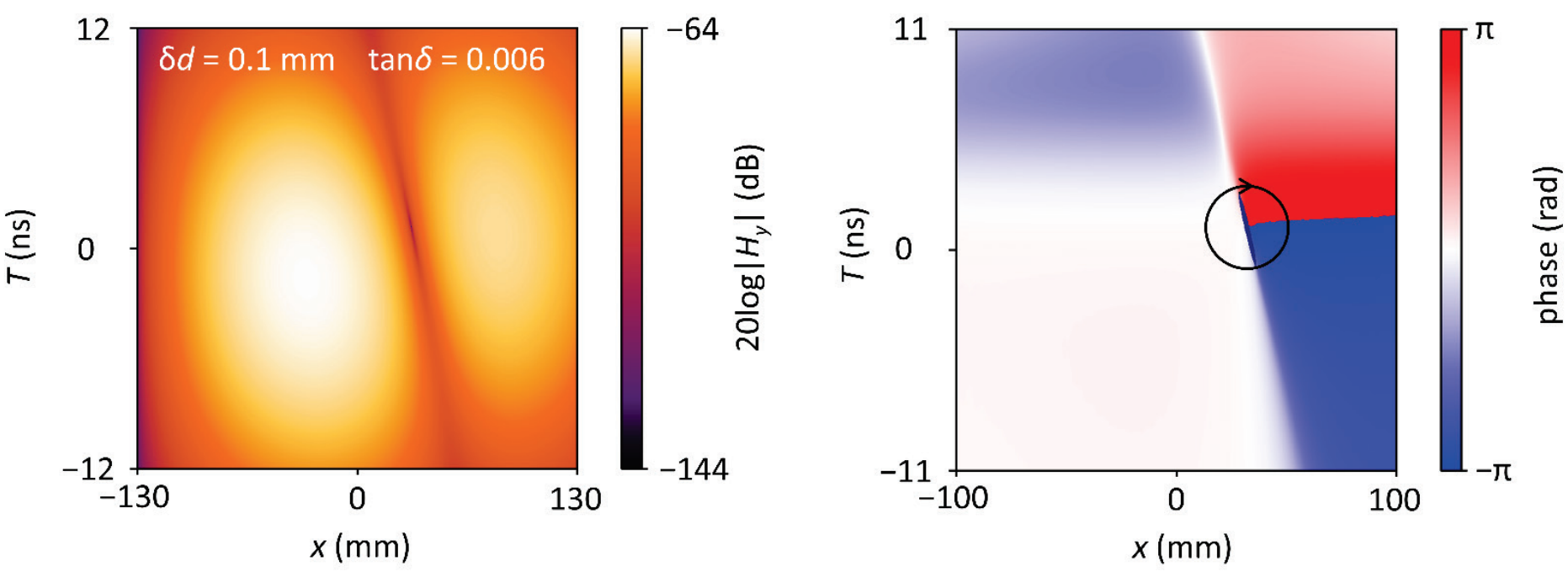

**Figure S12. Spatiotemporal response under oblique incidence excitation by a spatiotemporal Gaussian wave packet in the presence of material loss, with tan$\delta$ = 0.006.** The left panel shows the magnitude distribution of $H_y$ in the $x$ - $T$ domain. The right panel presents the corresponding phase distribution. Despite the introduction of loss, a transmission zero and an associated phase singularity persist, indicating the robustness of the STOV. Owing to material loss, the transmission zero shifts away from the Γ point toward positive $k_x$, where complete destructive interference is restored. This momentum-space shift is accompanied by a corresponding displacement of the phase singularity in the $x$ - $T$ domain. The accumulated phase winding around the singularity along a clockwise loop remains 2π, confirming a preserved winding number of −1.

Figure S12: Simulated space-time magnitude and phase maps show the transmitted field under oblique excitation with material loss. A shifted phase singularity remains present, confirming that the spatiotemporal optical vortex response is robust in a system with material loss.

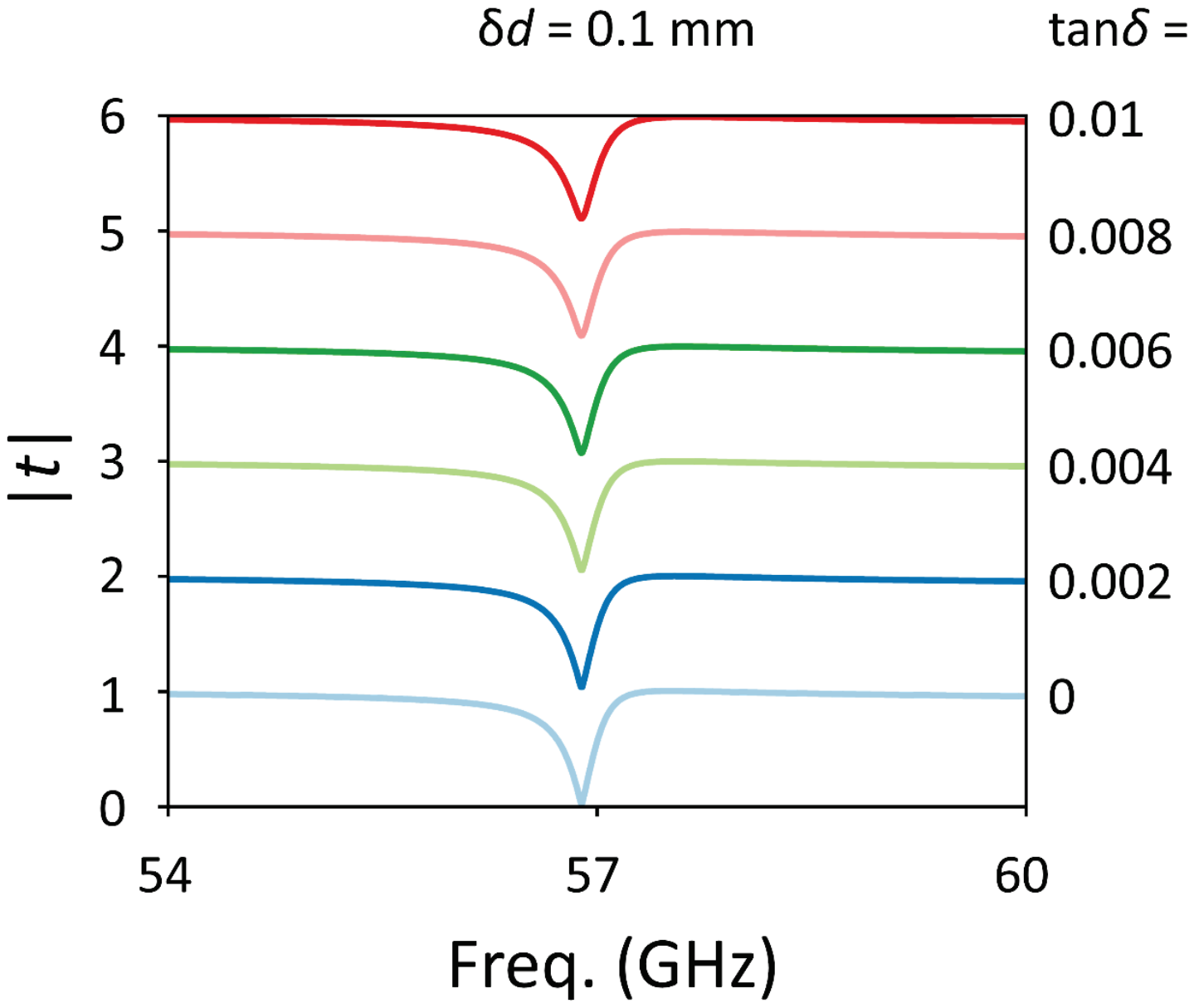

**Figure S13. Variation of the absolute transmission coefficient $|t|$ with tan$\delta$ for $\delta d$ = 0.1 mm.** The spectra are vertically offset by unity in sequence for clarity. Despite the gradual broadening of the resonance with increasing loss, the transmission dip remains close to zero over a wide range of the loss tangent tan$\delta$, indicating the persistence of the transmission zero under moderate material loss. This robustness under material loss further supports the qBIC picture underlying the observed angularly dispersive resonance.

Figure S13: Transmission spectra for different loss tangents show that the resonance broadens as material loss increases. The transmission dip remains close to zero over a moderate loss range, supporting the robustness of the qBIC mechanism.

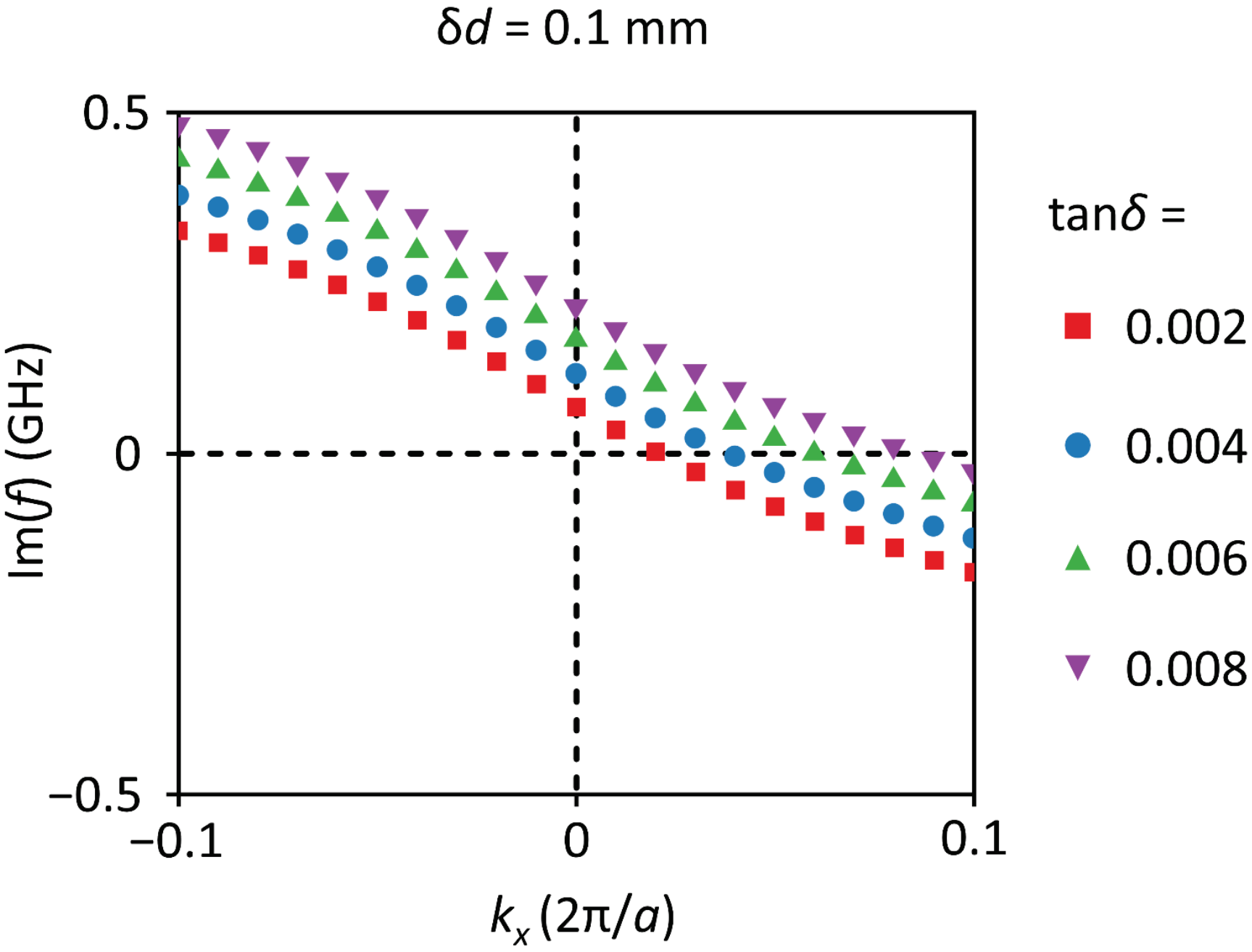

**Figure S14. Material-loss-induced shift of transmission-zero trajectories in the $k_x$ – Im($f$) space.** As tan$\delta$ increases, the trajectories of transmission zero shift toward larger |Im($f$)|, and the crossing with the $k_x$ axis moves toward positive $k_x$, indicating material-loss-induced modification of the resonance. Marker definitions: red squares, tan$\delta$ = 0.002; blue circles, tan$\delta$ = 0.004; green upward triangles, tan$\delta$ = 0.006; purple downward triangles, tan$\delta$ = 0.008.

Figure S14: Transmission-zero trajectories in momentum and complex-frequency space show the effect of increasing material loss. The zero shifts toward larger imaginary frequency values and positive momentum, indicating material-loss-induced modification of the resonance.

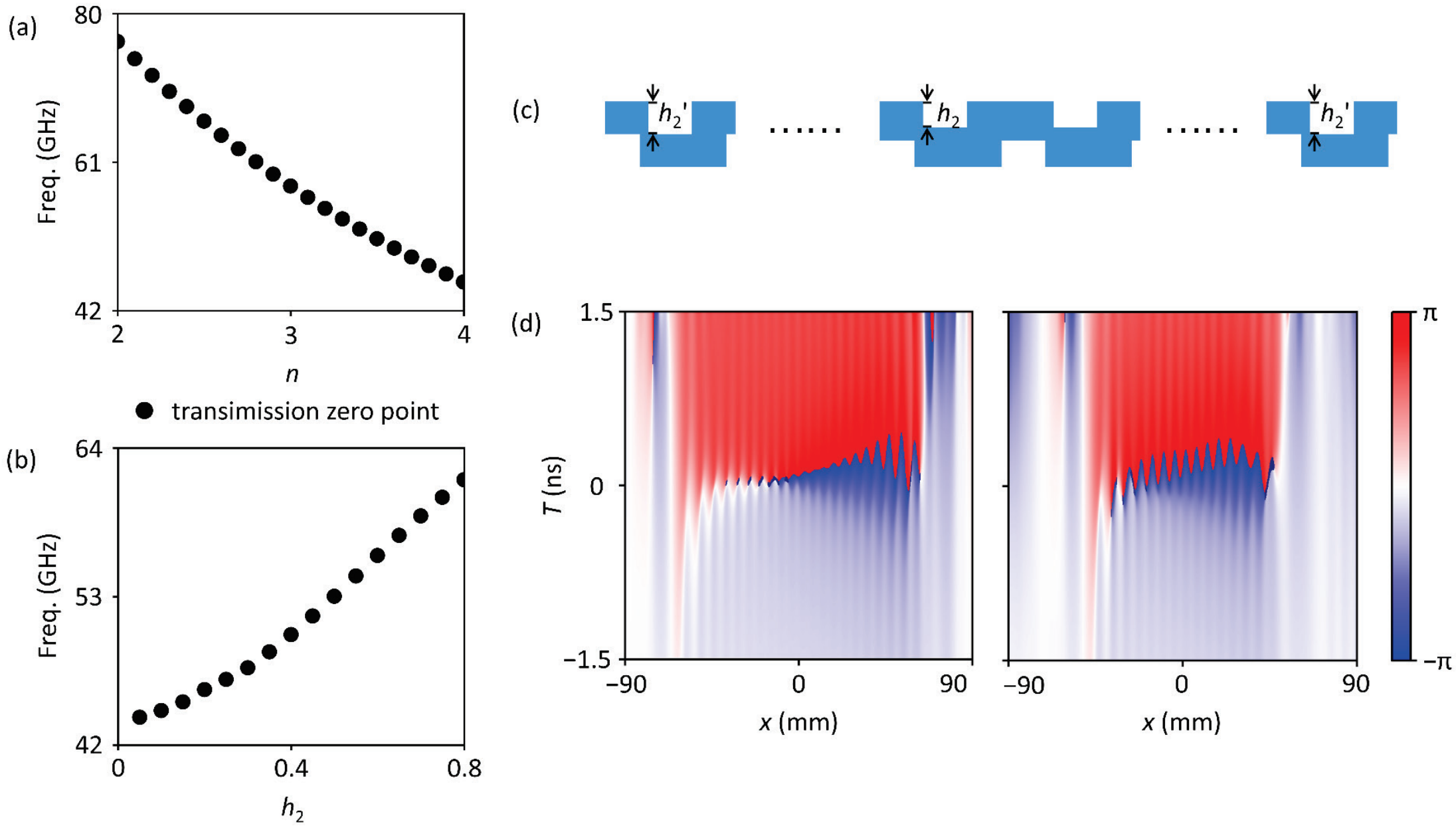

**Figure S15. Influence of material parameters and structural nonuniformity on the STOV response.** (a) Transmission-zero frequency as a function of refractive index $n$ at normal incidence with $\delta d$ = 0.3 mm. (b) Transmission-zero frequency as a function of the slot-depth-related parameter $h_2$, with $h_2 = h_3$. (c) Schematic of the finite-array model used to mimic fabrication-induced nonuniformity, where the central region has the nominal depth $h_2$ = 0.6 mm and the end regions have a larger depth $h_2'$ = 0.75 mm. (d) FDTD-simulated phase distributions in the $x$ - $T$ plane for 25 and 30 perturbed periods at each end, respectively. The results show that parameter mismatch shifts the transmission-zero frequency, while structural nonuniformity can distort the phase singularity and degrade the STOV profile.

Figure S15: Parameter and nonuniformity analyses show how refractive index, slot depth, and finite-array perturbations affect the vortex response. Material or structural deviations shift the transmission-zero frequency and can distort the phase singularity.

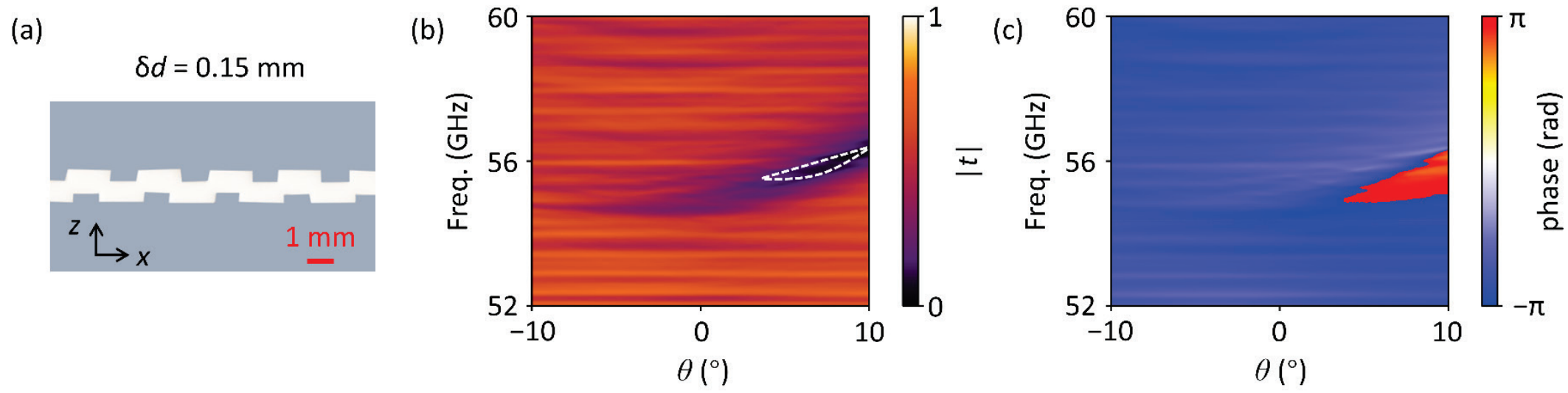

**Figure S16. Enhanced experimental observation of the branch-cut response using a sample with increased asymmetry parameter.** (a) Side-view optical image of the fabricated sample with an increased asymmetry parameter $\delta d$. The increased perturbation lowers the Q-factor of the qBIC resonance, thereby reducing its sensitivity to the finite number of periods in the experimental sample. (b) and (c) Measured transmission amplitude and phase distributions in the $\theta$ - $f$ space. Compared with the sample shown in the main text, the resonance linewidth becomes broader and the transmission dip is further reduced, with a minimum amplitude of approximately 0.015. The white dashed curves indicate the branch-cut region associated with the transmission zero. The corresponding phase response exhibits a more pronounced $2\pi$ coverage, confirming that increasing $\delta d$ facilitates a clearer experimental observation of the STOV-induced branch-cut feature.

Figure S16: Experimentally measured transmission amplitude and phase maps for a sample with increased asymmetry parameter modulation. The results show a broader resonance, a deeper transmission dip, and a clearer branch-cut-like phase feature that is less affected by finite sample size.